\documentclass[11pt,a4paper]{article}
\usepackage[T1]{fontenc}
\usepackage[utf8]{inputenc}

\newif\ifjhep
\IfFileExists{jheppub.sty}{\jheptrue}{\jhepfalse}
\ifjhep
  \usepackage{jheppub}
\else
  \usepackage[a4paper,margin=2.5cm]{geometry}
  \usepackage{amsmath,amssymb,graphicx,xcolor}
  \usepackage[numbers,sort&compress]{natbib}
  \usepackage[colorlinks=true,linkcolor=blue!55!black,citecolor=blue!55!black,urlcolor=blue!55!black]{hyperref}
  \makeatletter
  \def\jh@authors{}\def\jh@affils{}\def\jh@emails{}\def\jh@abstract{}\def\jh@keywords{}\def\jh@arxiv{}
  \renewcommand{\author}[2][]{\g@addto@macro\jh@authors{\mbox{#2\textsuperscript{\,#1}}\quad}}
  \newcommand{\affiliation}[2][]{\g@addto@macro\jh@affils{\textsuperscript{#1}\,\textit{#2}\par}}
  \newcommand{\emailAdd}[1]{\g@addto@macro\jh@emails{\texttt{#1}\quad}}
  \renewcommand{\abstract}[1]{\gdef\jh@abstract{#1}}
  \newcommand{\keywords}[1]{\gdef\jh@keywords{#1}}
  \newcommand{\arxivnumber}[1]{\gdef\jh@arxiv{#1}}
  \renewcommand{\maketitle}{\begingroup\thispagestyle{empty}\noindent
    {\LARGE\bfseries\raggedright\@title\par}\vspace{1.4em}
    \noindent{\large\raggedright\jh@authors\par}\vspace{0.8em}
    {\small\raggedright\jh@affils\par}\vspace{0.4em}
    \noindent{\small\textit{E-mail:}\ \jh@emails\par}\vspace{1.2em}
    \noindent\textsc{Abstract:} \jh@abstract\par\vspace{0.8em}
    \noindent{\small\textsc{Keywords:} \jh@keywords\par}\vspace{0.4em}
    \noindent{\small\textsc{ArXiv ePrint:} \jh@arxiv\par}\endgroup\clearpage}
  \newcommand{\acknowledgments}{\section*{Acknowledgments}}
  \numberwithin{equation}{section}
  \makeatother
\fi
\usepackage{amsmath,amssymb,graphicx,xcolor}
\usepackage{bm,mathtools,booktabs}
\allowdisplaybreaks

\newif\ifmarkup \markuptrue
\ifdefined\nomarkup\markupfalse\fi
\ifmarkup\else\fi
\ifmarkup\else\fi

\newcommand{\dd}{\mathrm{d}}
\newcommand{\mpl}{M_{\rm P}}
\newcommand{\Ms}{M_{\rm S}}
\newcommand{\Mnl}{M_{\rm nl}}
\newcommand{\Lstar}{\Lambda_*}
\newcommand{\GB}{\mathcal{G}}
\newcommand{\be}{\beta}
\newcommand{\Vst}{V_{\rm St}}
\newcommand{\VRG}{V_{\rm RG}}
\newcommand{\MSbar}{\overline{\rm MS}}
\newcommand{\ord}{\mathcal{O}}
\newcommand{\eps}{\epsilon}            
\newcommand{\epsd}{\varepsilon}        
\newcommand{\Boxop}{\Box}
\newcommand{\Trh}{T_{\rm rh}}
\newcommand{\Hf}{\mathcal{H}}           
\newcommand{\Npol}{\mathsf{N}}         
\newcommand{\Rc}{\mathcal{R}}          
\newcommand{\Amp}{\mathcal{A}}         
\newcommand{\hD}{\hat\Delta}           

\title{Removability criterion for radiative corrections to the Starobinsky attractor in weakly nonlocal gravity}

\author[a]{Ekapob Kulchoakrungsun,}
\author[b]{Phongpichit Channuie}
\author[a]{and Daris Samart}

\affiliation[a]{Khon Kaen Particle Physics and Cosmology Theory Group (KKPaCT), Department of Physics, Faculty of Science, Khon Kaen University, Khon Kaen 40002, Thailand}
\affiliation[b]{School of Science \& College of Graduate Studies, Walailak University, Nakhon Si Thammarat 80160, Thailand}

\emailAdd{ek2897@nyu.edu}
\emailAdd{phongpichit.ch@mail.wu.ac.th}
\emailAdd{darisa@kku.ac.th}

\abstract{In this work, we investigate how the one-loop running of local couplings in weakly nonlocal, super-renormalizable completions of $R+R^{2}$ gravity affects the predictions of the Starobinsky model. Our study is motivated by the larger scalar spectral index $n_s$ preferred by recent data. For a canonical inflaton at first order in slow roll, we derive the response of the observables to an arbitrary fractional deformation $\Delta$ of the Einstein-frame potential $V$. We find that it is governed by $\Rc[\Delta]=(V/V')\Delta'$, and that the shift of the logarithm of the slow-roll parameter is twice the logarithmic derivative of the trajectory average of $\Rc$. This provides a test which requires one differentiation only. Deformations with constant $\Rc$, which we call removable, change the E-model exponent and shift $n_s$ at $\ord(\ln N/N^{2})$ only, where $N$ is the number of $e$-folds, while $\Rc\propto N^{k}$ gives $\Delta n_s=[2k^{2}/(k+1)]\Rc/N$. When the running is improved at the Einstein-frame Hubble rate, the renormalization-group time is half the logarithm of the tree-level potential and the gravitational flow is removable. Under a naturalness condition, a matching to the nonlocality scale yields $|\Delta n_s|\lesssim3\times10^{-4}$, a factor of about $30$ below the shift preferred by the Atacama Cosmology Telescope (ACT). However, for a scale tied to the Jordan-frame curvature, the same flow is not removable and $\Delta n_s\simeq K/3$, where $K$ is a linear combination of the running parameters. Removability is a property of the flow together with the scale prescription. None of the deformations proposed since ACT DR6 that raise $n_s$ to the preferred value is removable. These results hold within a projected truncation of the local couplings and a first-order improvement of the potential.}

\keywords{Models of Quantum Gravity, Renormalization Group, Cosmology of Theories BSM, Classical Theories of Gravity}
\arxivnumber{2608.17215}

\begin{document}
\maketitle
\flushbottom
\tableofcontents

\section{Introduction}
\label{sec:intro}

The Starobinsky model \cite{Starobinsky1980} is the lowest member of the higher-derivative family whose renormalization properties were established in Ref.~\cite{Stelle1977}. It predicts $n_s\simeq1-2/N$ and $r\simeq12/N^{2}$, where $n_s$ is the scalar spectral index, $r$ is the tensor-to-scalar ratio and $N$ is the number of $e$-folds before the end of inflation, with no freedom beyond the scalaron mass, which is fixed by the measured scalar amplitude. Recently, the combination of the Atacama Cosmology Telescope (ACT) Data Release 6 (DR6) with Planck, cosmic microwave background (CMB) lensing, and the Dark Energy Spectroscopic Instrument (DESI) DR1 baryon acoustic oscillation (BAO) data gives $n_s=0.9743\pm0.0034$ \cite{ACTDR6,ACTDR6ext}, which lies $2.7\sigma$ above the tree-level value $0.9650$ at $N_*=55$ and $1.9\sigma$ above it at $N_*=60$ \cite{StaroACTtension}, where $N_*$ denotes the value of $N$ at which the pivot scale leaves the horizon, while the tensor bound remains compatible \cite{Planck2018Inflation,BICEPKeck2021}. However, it is worth noting that the size of this preference depends on the data combination, in particular on the inclusion of the BAO data, whose consistency with the CMB within $\Lambda$CDM is under discussion \cite{SPT3GD1,FerreiraBAO}. Indeed, a combined analysis finds $n_s=0.9682\pm0.0032$ from the CMB data alone (Planck, the South Pole Telescope (SPT), ACT and BICEP/Keck) and $n_s=0.9728\pm0.0029$ when the DESI BAO data are added \cite{Balkenhol2025}. The tree-level value at $N_*=55$ lies $1\sigma$ below the former and $2.7\sigma$ below the latter. Moreover, the values of $n_s$ and of its running inferred from the CMB depend on the assumed pre-recombination expansion history \cite{Sabogal:2026qvy}. In this work, we take the displacement $\Delta n_s=+9.3\times10^{-3}$ between the ACT central value and the tree-level prediction as a reference scale, without assuming that it will persist.

There are several proposals to solve this problem, e.g., higher-curvature operators \cite{CurvACT}, revised post-inflationary histories \cite{StaroACTrefined,Iacconi:2025odq}, and radiative corrections. In particular, it was shown in Ref.~\cite{AttractorsRadACT} that percent-level and even sub-percent-level radiative corrections move the predictions of $\alpha$- and $\xi$-attractors into the favored region. Moreover, the authors of Ref.~\cite{Ellis2025} found that Yukawa and trilinear couplings of the scalaron shift $n_s$ enough for Planck and ACT to constrain them. For $\alpha$-attractors, it was shown in Ref.~\cite{Iacconi:2025odq} that a stiff reheating phase raises $n_s$ up to $0.9682$ for monomial T-models, while a binomial potential leads to a non-universal behavior of $n_s$ \cite{Iacconi:2026twl}. With corrections that small, the predictions are only as sharp as the ultraviolet (UV) control over them. We therefore address this question in weakly nonlocal, super-renormalizable, or finite quantum gravity \cite{Krasnikov1987,Kuzmin1989,Tomboulis1997,ModestoRachwal2014,Briscese2013,Modesto2013,Koshelev2016,KumarModesto2018,KoshelevKumarStarobinsky2023,ModestoOrlando2026}, in which the Starobinsky solution is embedded exactly.

Before proceeding further, it is worth noting that the word ``nonlocal'' carries two distinct meanings in the literature. Here it denotes the UV nonlocality, where the action contains entire functions of the d'Alembertian $\Boxop$ devoid of zeros. Such a form factor damps the propagator at large momenta without adding states to the spectrum, and this is what guarantees super-renormalizability. This framework follows the Krasnikov, Kuz'min, Tomboulis, and Modesto construction \cite{Krasnikov1987,Kuzmin1989,Tomboulis1997,ModestoRachwal2014}. On the other hand, the infrared (IR) nonlocality describes theories containing $\Boxop^{-1}$ or functions of $\Boxop^{-1}R$ \cite{DeserWoodard2007,NojiriOdintsovOikonomou2020,NojiriOdintsovOikonomou2026}. Those theories are built to describe the late-time acceleration. They are not super-renormalizable, and the one-loop exactness used below does not apply to them. In addition, an IR-nonlocal deformation depends on the history of the background through $\Boxop^{-1}R$ and is not a function of the instantaneous potential alone. Therefore, the test formulated in this paper does not apply to it.

The present work builds on and extends the results in the literature, and it is worth recalling them before presenting our study. Firstly, the Starobinsky solution survives exactly in a weakly nonlocal completion, and hence the tree-level inflationary predictions are inherited. This framework is detailed in Refs.~\cite{Briscese2013,Modesto2013,Koshelev2016} and generalized in Ref.~\cite{KoshelevKumarStarobinsky2023}, while the non-Gaussianities and the observational reach were studied in Refs.~\cite{KoshelevKumarMazumdarStarobinsky2020,NonGauss2022} and Ref.~\cite{KumarModesto2018}, respectively. The Einstein-frame completion quadratic in the equations of motion and the observation that loop-induced propagator poles lie outside the perturbative regime are due to Ref.~\cite{ModestoOrlando2026}. That work states that the classical predictions survive up to perturbative corrections without computing them, and those corrections are the subject of the present paper. Secondly, the divergent local beta functions of super-renormalizable completions are one-loop exact \cite{ModestoRachwal2014,ModestoRachwalShapiro2018,RachwalModestoPinzulShapiro2021}, and this property was extended to gravity nonminimally coupled to matter in Ref.~\cite{Calcagni2023}. The coefficients used below are taken from Ref.~\cite{ModestoRachwalShapiro2018}, with the sign conventions of Ref.~\cite{ModestoOrlando2026}. Thirdly, the observables of a single-field slow-roll model are functionals of the slow-roll parameter $\eps(N)$ only, and $\eps(N)$ determines the model only up to integration constants. This relies on the $N$-formalism of Refs.~\cite{Mukhanov2013,Roest2014,GarciaBellidoRoest2014}, where the exponent independence of the E-model predictions is a standard property of cosmological $\alpha$-attractors \cite{KalloshLindeRoest2013,KalloshLindeRoest2014,Roest2014}, including the logarithmic extensions of Ref.~\cite{Linder2021}. In the pole-inflation formulation, the predictions are fixed by the leading behavior of the kinetic function near its pole, and it was analyzed in Ref.~\cite{BroyGalanteRoestWestphal2015} which corrections to that structure preserve them \cite{GalanteKalloshLindeRoest2015}. The criterion derived below is the statement of that property for a multiplicative deformation of the canonical Einstein-frame potential. Fourthly, the logarithmic running of the coefficient of $R^{2}$ has been studied before. With the renormalization scale tied to the Jordan-frame curvature, it produces an $R^{2}\ln R$ term, or equivalently a marginal deformation $R^{2(1-\alpha)}$, and the spectral index shifts at first order in the running coefficient \cite{BenDayan2014,CodelloEtAl2015,LiuProkopecStarobinsky2018}. Note that the dependence of radiative corrections on the frame in which the scale is defined is also familiar from Higgs inflation \cite{BezrukovShaposhnikov2009,GeorgeMooijPostma2014,KamenshchikSteinwachs2015}.

The main results of this work are summarized as follows:
\begin{itemize}
\item[(i)] \textbf{A response formula for an arbitrary deformation.} We derive the first-order response of $n_s$ and $r$ to a fractional deformation $\Delta$ of the Starobinsky potential, see Eq.~\eqref{eq:response}. We find that it depends on $\Delta$ through $\Rc[\Delta]=(\Vst/\Vst')\Delta'$ only and that, on the plateau, its leading term is the logarithmic derivative of the trajectory average of $\Rc$ (Proposition~1 of Sec.~\ref{sec:iff}). The formula holds for every differentiable deformation, includes the displacement of the end point, and agrees with the numerical solution to a relative accuracy of $10^{-6}$ in the linear regime.
\item[(ii)] \textbf{A test that requires one differentiation.} The response vanishes identically when $\Rc$ is constant. We call such deformations removable, because they change the E-model exponent, which is absorbed at leading order by an amplitude rescaling and a shift of the inflaton. Moreover, the behavior of $\Rc$ on the plateau sorts deformations into three classes with distinct scalings, and $\Rc\propto N^{k}$ gives $\Delta n_s=[2k^{2}/(k+1)]\Rc/N$. The attractor literature states which structures preserve the leading predictions. Here we add the size of the shift when they are not preserved, without a computation of observables.
\item[(iii)] \textbf{Gravitational running at the Einstein-frame scale.} For a completion defined in the Einstein frame and improved at the Einstein-frame Hubble rate, we show that the renormalization group (RG) time is one half of $\ln\Vst$, and therefore the one-loop gravitational running is removable. We obtain its residual effect on $n_s$, which is $\ord(\ln N/N^{2})$, in closed form for the constant-$\Rc$ direction and at leading-logarithmic order for all channels. In addition, we provide the frame-consistent treatment of the running Planck mass, the variation of the matching of the pivot scale, and a conditional estimate $|\Delta n_s|\lesssim3\times10^{-4}$.
\item[(iv)] \textbf{One rule for the scale dependence.} An improvement scale that varies as $u^{-s}$ on the plateau gives $\Delta n_s\simeq\tfrac23sK$, where $u\equiv e^{-\sqrt{2/3}\,\phi/\mpl}$, $\phi$ is the canonical inflaton field, $\mpl$ is the reduced Planck mass and $K$ is the combination of running parameters defined in Eq.~\eqref{eq:gradients}. The Einstein-frame Hubble rate has $s=0$, while the Jordan-frame curvature has $s=\tfrac12$ and reproduces the marginal deformation of Refs.~\cite{BenDayan2014,CodelloEtAl2015,LiuProkopecStarobinsky2018}, which is not removable. Removability is therefore a property of the flow together with the scale prescription, and we state the prescription as an assumption.
\item[(v)] \textbf{Classification of the proposed deformations since ACT DR6.} The test separates them by a single differentiation. In particular, matter-induced corrections with field logarithms are not removable, and this is why they shift the predictions \cite{AttractorsRadACT,Ellis2025}.
\end{itemize}
To our knowledge, the response formula in Eq.~\eqref{eq:response} and the general rule for the scale dependence in item (iv) have not been obtained previously. The residual of the E-model direction in item (iii) extends the known large-$N$ expansion \cite{Roest2014,GarciaBellidoRoest2014,Linder2021} to the running parameters, including the vacuum-energy and Planck-mass channels. It is worth noting that the results (i) and (ii) are statements about canonical plateau inflation and are independent of any completion. On the other hand, the results (iii) and (iv) are conditional. They rest on a projected truncation of the local couplings, on a first-order improvement of the potential, and on the identification of the scale. Their translation into predictions of a particular UV completion requires the matching and effective-action assumptions stated in Secs.~\ref{sec:completion} and \ref{sec:limits}.

This paper is organized as follows. In Sec.~\ref{sec:setup}, we fix the conventions and derive the response of the plateau predictions to a deformation, without reference to a completion. In Sec.~\ref{sec:completion}, we specify the completions, the one-loop running, the improved potential, and the frame normalization. We then apply the criterion to the gravitational flow and quantify the residual channels in Sec.~\ref{sec:results}. The perturbative estimates for the propagator and the tensor sector are collected in Sec.~\ref{sec:poles}. In Sec.~\ref{sec:discussion}, we discuss matter-induced corrections, the deformations proposed since ACT DR6, the improvement scale, and the limitations. Finally, we summarize our conclusions in Sec.~\ref{sec:concl}.

\section{Plateau inflation and its response to a deformation}
\label{sec:setup}

In this section, we fix the normalization and derive the first-order response of the slow-roll predictions to a deformation of the potential. No property of a UV completion is used here.

\subsection{Conventions and the Starobinsky potential}
\label{sec:frames}

Throughout this work, we adopt the metric signature $(-,+,+,+)$ and the Riemann tensor convention $R^{\rho}{}_{\sigma\mu\nu}=\partial_\mu\Gamma^{\rho}_{\nu\sigma}-\partial_\nu\Gamma^{\rho}_{\mu\sigma}+\Gamma^{\rho}_{\mu\lambda}\Gamma^{\lambda}_{\nu\sigma}-\Gamma^{\rho}_{\nu\lambda}\Gamma^{\lambda}_{\mu\sigma}$, $R_{\mu\nu}=R^{\rho}_{\ \mu\rho\nu}$, so that de Sitter space has $R=12H^{2}>0$. The reduced Planck mass $\mpl^{2}=1/(8\pi G)$, with $G$ the Newton constant, is kept explicit, and $\Boxop\equiv g^{\mu\nu}\nabla_\mu\nabla_\nu$ denotes the d'Alembertian. Loops are regularized in $D=4-2\epsd$ dimensions in the modified minimal subtraction scheme $1/\bar\epsd=1/\epsd-\gamma_E+\ln4\pi$, with the RG time $t=\ln(\mu/\mu_0)$, where $\gamma_E$ is the Euler--Mascheroni constant, $\mu$ is the renormalization scale and $\mu_0$ is a reference scale. Here the symbol $\eps$ is reserved for the slow-roll parameter, $H$ for the Hubble rate, and $N$ for the number of $e$-folds. In addition, the exponent of the form factor is written as $\Hf$ and the degree of its asymptotic polynomial as $\Npol$. The curvature-squared operators obey the following identity,
\begin{equation}
C^2=2R_{\mu\nu}^2-\tfrac23R^2+\GB \,,
\label{eq:C2identity}
\end{equation}
where $C^{2}\equiv C_{\mu\nu\rho\sigma}C^{\mu\nu\rho\sigma}$ is the square of the Weyl tensor, $R_{\mu\nu}^{2}\equiv R_{\mu\nu}R^{\mu\nu}$, and $\GB\equiv R_{\mu\nu\rho\sigma}R^{\mu\nu\rho\sigma}-4R_{\mu\nu}^{2}+R^{2}$ is the Gauss--Bonnet density. We use this identity to simplify $C^{2}$ for transverse-traceless perturbations. For a scalar field, we take the Lagrangian density $\mathcal{L}_\phi=-\frac12(\nabla\phi)^{2}-V-\frac12\xi R\phi^{2}$, with the nonminimal coupling $\xi$ and the effective mass squared $m_{\rm eff}^{2}=V''+\xi R$. Throughout this work, a prime on a function of the field denotes a derivative with respect to $\phi$, whereas $f'\equiv\dd f/\dd R$ for a function of the curvature.

We start with the Jordan-frame action $S_J=\int\dd^4x\sqrt{-g}\,\tfrac12\mpl^{2}f(R)$ with $f(R)=R+R^{2}/6\Ms^{2}$, where $\Ms$ is the scalaron mass. We then perform the conformal transformation $g_{\mu\nu}^E=f'(R)g_{\mu\nu}^J$ together with the field redefinition $\phi=\sqrt{3/2}\,\mpl\ln f'$. As a result, the Einstein-frame potential takes the form
\begin{equation}
\Vst(\phi)=V_0\left(1-u\right)^{2},
\qquad
V_0\equiv\frac34\mpl^{2}\Ms^{2},
\qquad
u\equiv e^{-a\phi},
\qquad
a\equiv\sqrt{\tfrac23}\,\frac{1}{\mpl}.
\label{eq:Vst}
\end{equation}
The relevant derivatives, the slow-roll parameters, the mapping to the number of $e$-folds, and the boundary condition $u_{\rm end}=2\sqrt3-3=0.4641$ are collected in Appendix~\ref{app:derivs}. We work at first order in slow roll, with $n_s=1-6\eps_V+2\eta_V$, $r=16\eps_V$, and $A_s=V/(24\pi^{2}\eps_V\mpl^{4})$,
where $\eps_V\equiv\tfrac12\mpl^{2}(V'/V)^{2}$ and $\eta_V\equiv\mpl^{2}V''/V$ are the potential slow-roll parameters.
The higher-order spectral terms \cite{StewartLyth1993} and the exact background evolution are examined in Sec.~\ref{sec:limits}. Here $N_*$ denotes the number of $e$-folds before the end of inflation at which the pivot scale leaves the horizon, and $\phi_0\equiv\phi_{N_*}$ is the corresponding field value. More generally, $\phi_N$ and $u_N\equiv u(\phi_N)$ denote the values reached $N$ $e$-folds before the end of inflation. At $N_*=55$, the measured scalar amplitude $A_s=2.1\times10^{-9}$ fixes
\begin{equation}
\begin{aligned}
\Ms&{}=1.22\times10^{-5}\mpl=2.97\times10^{13}\ {\rm GeV},\\
n_s&=0.96498,
\qquad
r=3.498\times10^{-3},
\qquad
u_0=1.264\times10^{-2},
\end{aligned}
\label{eq:Msfix}
\end{equation}
where $u_0\equiv u_{N_*}$ and we use $\mpl=2.435\times10^{18}$~GeV. The corresponding Hubble rate is $H_*=\tfrac12\Ms(1-u_0)=6.02\times10^{-6}\mpl=1.47\times10^{13}$~GeV. The leading-order relation $A_s=N_*^{2}\Ms^{2}/24\pi^{2}\mpl^{2}$ would give $\Ms=1.28\times10^{-5}\mpl$, which is five percent higher. The shifts of the observables are differences and are computed reliably at this order, whereas the absolute value of $n_s$ is uncertain at $\ord(10^{-4})$ from second-order slow-roll terms. In addition, the exact background evolution with the second-order spectral formula gives $n_s=0.96489$ and $r=3.550\times10^{-3}$ at the same $N_*$. These are the tree-level values that the completion inherits, and every shift computed in Sec.~\ref{sec:results} is measured against them.

\subsection{First-order response and removable deformations}
\label{sec:iff}

The statements of this subsection hold for a canonical scalar field with fixed $\mpl$ in Einstein gravity, on a monotonic branch of its potential and at first order in slow roll. In the slow-roll $N$-formalism, $N$ is the number of $e$-folds remaining until the end of inflation, so that $\eps=\dd\ln H/\dd N$. At first order in slow roll, $\eps$ coincides with the potential slow-roll parameter $\eps_V$ and with the Hubble slow-roll parameter $\eps_H\equiv-\dot H/H^{2}$, where an overdot denotes a derivative with respect to cosmic time. The observables are given by
\begin{equation}
n_s-1=-2\eps+\frac{\dd\ln\eps}{\dd N}\,,
\qquad
r=16\eps \,,
\label{eq:Nformalism}
\end{equation}
and hence both are functionals of $\eps(N)$ alone. Conversely, $\eps(N)$ reconstructs the background dynamics. Integrating $\dd\ln H/\dd N=\eps$ determines $H$ up to a single multiplicative constant, and $\dd\phi/\dd N=\mpl\sqrt{2\eps}$ determines $\phi$ up to a single additive constant. Therefore, two single-field slow-roll models have the same $\eps(N)$ on an interval, and hence identical $n_s$ and $r$ there, if and only if their potentials differ by a constant amplitude rescaling combined with a constant inflaton shift, up to the orientation $\phi\to-\phi$. Note that the reconstruction extends beyond slow roll when $\eps$ is the Hubble slow-roll parameter, since $V=\mpl^{2}H^{2}(3-\eps_H)$. We call such potentials \emph{exactly equivalent}. At the infinitesimal level, the exactly equivalent deformations of $\Vst$ read
\begin{equation}
\Delta_{\rm exact}=c_1+c_3\,\frac{\Vst'}{\Vst} ,
\label{eq:exactclass}
\end{equation}
where $c_1$ and $c_3$ are infinitesimal constants which generate the amplitude rescaling and the inflaton shift, respectively. The parameter $c_2$ is defined for the direction introduced in Eq.~\eqref{eq:Rtest}.
However, a larger class preserves the leading-order predictions. To identify this, we compute the response of $\eps(N)$ to a deformation $V=\Vst[1+\Delta]$ at first order in $\Delta$, where $\Delta(\phi)$ is a dimensionless function of the field with $|\Delta|\ll1$. The derivation is given in Appendix~\ref{app:response}, and here we briefly repeat its steps. We first define
\begin{equation}
\Rc[\Delta]\equiv\frac{\Vst}{\Vst'}\,\Delta' ,
\label{eq:Rdef}
\end{equation}
which we call the removability functional. It measures the fractional change of the logarithmic slope of the potential.
One then finds $(\ln V)'=(\ln\Vst)'\,[1+\Rc]$, so that at fixed field $\eps=\eps_0(1+2\Rc)$ and $\dd N=\dd N_0\,(1-\Rc)$, where the subscript $0$ denotes the tree-level model. Note that this applies to $\eps_0$ and $N_0$ only, whereas $u_0$, $\phi_0$ and $\mu_0$ denote quantities evaluated at the horizon exit of the pivot scale. The field value reached $N$ $e$-folds before the end therefore corresponds to the tree-level $e$-fold number $N+I(N)$, with $I(N)=\int_0^{N}\Rc\,\dd N'+u_{\rm end}\Rc_{\rm end}$, where the last term is the displacement of the end point. Evaluating $\eps$ there, we obtain
\begin{equation}
\begin{aligned}
\delta\ln\eps(N)&=2\,\Rc(N)+\frac{\dd\ln\eps_0}{\dd N}
\left[\int_0^{N}\Rc(N')\,\dd N'+u_{\rm end}\Rc_{\rm end}\right],\\
\frac{\Delta r}{r}&=\delta\ln\eps,
\qquad
\Delta n_s=-2\eps_0\,\delta\ln\eps+\frac{\dd}{\dd N}\delta\ln\eps ,
\end{aligned}
\label{eq:response}
\end{equation}
where $\delta\ln\eps(N)\equiv\ln\eps(N)-\ln\eps_0(N)$, $\Delta n_s$ and $\Delta r$ are the shifts of the observables with respect to the tree-level model at the same $N$, $\Rc_{\rm end}\equiv\Rc(N=0)$ is the value at the end point of the tree-level trajectory, $\Rc(N)$ is $\Rc$ evaluated on the tree-level trajectory and the derivative in $\Delta n_s$ is taken at fixed $\phi_0$ since $\Delta$ may depend on $\phi_0$ through the improvement scale, see Eq.~\eqref{eq:ell}. It is worth noting that Eq.~\eqref{eq:response} is exact at first order in $\Delta$ and first order in slow roll. Moreover, it reproduces the numerical solution of Appendix~\ref{app:num} in every channel considered below to the accuracy of the linearization. For the Starobinsky plateau, i.e., the region $u\ll1$, or equivalently $N\gg1$ since $u_N\simeq3/4N$, one has $\eps_0=3/4N^{2}$ at leading order, and Eq.~\eqref{eq:response} becomes
\begin{equation}
\delta\ln\eps(N)=2\left[\Rc(N)-\langle\Rc\rangle_N\right]
=2\,\frac{\dd\langle\Rc\rangle_N}{\dd\ln N},
\qquad
\langle\Rc\rangle_N\equiv\frac1N\int_0^{N}\Rc(N')\,\dd N' ,
\label{eq:average}
\end{equation}
up to the end-point term, which is $\ord(1/N)$. This means that the shift of $\ln\eps$ is twice the logarithmic derivative of the trajectory average of $\Rc$. It vanishes identically at this order if and only if $\Rc$ is constant. Here the cancellation between the local term $2\Rc$ and the displacement of the $e$-fold map relies on $\eps_0\propto N^{-2}$. For $\eps_0\propto N^{-b}$ with a constant exponent $b$, a constant $\Rc=\bar\Rc$ gives $\delta\ln\eps=(2-b)\bar\Rc$,
and therefore the criterion is specific to the universality class of the plateau models with an exponential approach \cite{Roest2014}. We collect these statements as follows.

\medskip
\noindent\textbf{Proposition 1.} Let $V=\Vst[1+\Delta]$, with $\Delta$ differentiable on the inflationary trajectory and $\Rc(N)$ locally integrable, and work at first order in $\Delta$ and in slow roll.
\begin{itemize}
\item[(a)] Eq.~\eqref{eq:response} holds for all $N$.
\item[(b)] For $N\gg1$, $\delta\ln\eps=2\,\dd\langle\Rc\rangle_N/\dd\ln N$, up to the end-point term $-2u_{\rm end}\Rc_{\rm end}/N$ and to corrections of relative order $\ln N/N$.
\item[(c)] The leading prediction for $r$ is preserved, in the sense that $\Delta r/r\to0$ as $N\to\infty$, if and only if the derivative $\dd\langle\Rc\rangle_N/\dd\ln N$ tends to zero. It is sufficient that $\Rc(N)$ tends to a constant. If $\Rc$ has power-law or logarithmic asymptotics, then also $N\Delta n_s\to0$.
\item[(d)] If $\Rc\propto N^{k}$ with $k>-1$ and $k\neq0$, then at leading order
\begin{equation}
\delta\ln\eps=\frac{2k}{k+1}\,\Rc(N),
\qquad
\Delta n_s=\frac{2k^{2}}{k+1}\,\frac{\Rc(N)}{N} .
\label{eq:nonremovable}
\end{equation}
\end{itemize}
Part (a) is derived in Appendix~\ref{app:response}. Part (b) follows from $\eps_0\simeq3/4N^{2}$, whose corrections are of relative order $\ln N/N$, and parts (c) and (d) follow from (b). We call deformations that satisfy the condition of part (c) \emph{leading-order equivalent} to the Starobinsky potential. The exact constancy of $\Rc$ is sufficient for the leading-order equivalence and is not necessary.

The condition of exact constancy reads
\begin{equation}
\Rc[\Delta]={\rm const}
\qquad\Longleftrightarrow\qquad
\Delta=c_1+c_2\ln\frac{\Vst}{A_*} ,
\label{eq:Rtest}
\end{equation}
which is equivalent to $V\to A_*(V/A_*)^{1+c_2}$ on the plateau with $A_*$ being a fixed normalization with the dimension of an energy density. We designate such deformations as \emph{removable}, and refer to $c_2$ as the logarithmic E-model direction. Expanding the potential, we find $A_*(\Vst/A_*)^{1+c_2}\propto1-(2+2c_2)u+\ord(u^{2})$. A removable deformation therefore rescales the overall amplitude, which is absorbed by $A_s$, and rescales the coefficient of $e^{-a\phi}$, which is absorbed by the field redefinition,
\begin{equation}
\tilde\phi=\phi-\frac{\ln(1+c_2)}{a}
\label{eq:shift}
\end{equation}
leaving the kinetic term canonical. In other words, a removable deformation leaves the leading approach to the plateau, $V\propto1-c\,e^{-a\phi}$ intact with an unchanged $a$ and a positive constant $c$. This is the form in which the universality of the attractor predictions is usually stated \cite{KalloshLindeRoest2013,GalanteKalloshLindeRoest2015,BroyGalanteRoestWestphal2015}. More importantly, Eq.~\eqref{eq:Rtest} turns it into a test that requires one differentiation.

Eqs.~\eqref{eq:exactclass} and \eqref{eq:Rtest} characterize distinct classes. Their intersection is the constant $c_1$ only, and neither class contains the other. Together they span the combined space,
\begin{equation}
\Delta_{\rm inv}=c_1+c_2\ln\frac{\Vst}{A_*}+c_3\,\frac{\Vst'}{\Vst} ,
\label{eq:invclass}
\end{equation}
which corresponds to the tangent space of the E-model family $V=A(1-e^{-a(\phi-\phi_{\rm s})})^{p}$ at $p=2$. Here $A$, $p$ and $\phi_{\rm s}$ are the amplitude, the exponent and the origin of the field, respectively, and the three independent directions are $\partial_A$, $\partial_p$, and $\partial_{\phi_{\rm s}}$. The predictions of this family are exponent independent at leading order, and hence every direction in Eq.~\eqref{eq:invclass} is invisible at that order. This three-parameter family is a subfamily of the leading-order equivalent deformations, and $\Delta=c\,u^{q}$ with constants $c$ and $q>1$ is an example outside it. Evaluating the criterion on the combined space, one obtains
\begin{equation}
\Rc\big[\Delta_{\rm inv}\big]=c_2+c_3\,\mathcal{Q},
\qquad
\mathcal{Q}\equiv\frac{\Vst}{\Vst'}\left(\frac{\Vst'}{\Vst}\right)'
=\frac{\Vst''}{\Vst'}-\frac{\Vst'}{\Vst}
=-\frac{a}{1-u} .
\label{eq:Gfun}
\end{equation}
This shows that $\Rc$ is constant throughout the class up to the variation of $\mathcal{Q}$, which is $\ord(u)$ on the plateau. The field-shift direction is therefore not excluded by Eq.~\eqref{eq:Rtest}. It satisfies the criterion to $\ord(u)=\ord(1/N)$, i.e., the accuracy at which $\ln\Vst$ and $\Vst'/\Vst$ become degenerate, both being proportional to $u$ at leading order.

These observations fix the operational reading of the criterion, which we use throughout this work. On the plateau, a deformation falls into one of three classes according to the behavior of $\Rc$ as $u\to0$:
\begin{itemize}
\item[R.] \emph{Removable at leading order}: $\Rc=c_2+\ord(u)$ with $c_2\neq0$. By using Eq.~\eqref{eq:average}, the constant cancels, and both the removable residual and the $\ord(u)$ variation of $\Rc$ contribute $\Delta n_s=\ord(c_2\ln N/N^{2})$ and $\Delta r/r=\ord(c_2\ln N/N)$. The coefficient of $\ln N/N^{2}$ depends on the $\ord(u)$ term and is not universal.
\item[S.] \emph{Subleading}: $\Rc\to0$. The deformation is leading-order equivalent by part (c) of Proposition~1, and the size of the shift depends on how fast $\Rc$ decays. For $\Rc=\ord(u\ln u)$ or smaller, it is of the same order as in class~R, with at most an additional power of $\ln N$. For a slower decay, it is larger. For instance, the deformation $\Delta=c\,u^{q}$ with $1<q<2$ has $\Rc=-\tfrac12cq(1-u)u^{q-1}\propto N^{1-q}$, and Eq.~\eqref{eq:nonremovable} with $k=1-q$ gives $\Delta n_s\propto N^{-q}$, which lies between the tree-level tilt and the class-R residual. However, this asymptotic form is approached slowly, because it competes with terms of order $1/N^{2}$. For $q=\tfrac32$, the ratio of the first-order shift to Eq.~\eqref{eq:nonremovable} is $0.14$ at $N=55$, $0.73$ at $N=10^{3}$ and $0.97$ at $N=10^{5}$.
\item[N.] \emph{Not removable}: $\Rc$ grows without bound on the plateau. For $\Rc\propto N^{k}$ with $k>0$, Eq.~\eqref{eq:nonremovable} gives a shift larger than the class-R shift by a factor of order $N\,\Rc/(c_2\ln N)$, and the leading predictions change.
\end{itemize}
In summary, the exact constancy of $\Rc$ isolates the direction $c_2$, the constancy at $\ord(u)$ selects class~R, and the leading-order equivalence comprises classes R and S. We will see in Sec.~\ref{sec:classification} that two diagnostics calibrate this reading.

\section{Weakly nonlocal completions and one-loop running}
\label{sec:completion}

The proposition of Sec.~\ref{sec:iff} needs a deformation as input, and in this section we specify how it is obtained. It is worth noting that we do not derive the beta functions, the scalar--metric Hessian, and the perturbation spectra from a single gauge-fixed action. Our analysis does not rely on a specific completion in the following sense. The running parameters $\theta_i=(\gamma_M,b_G,\hat B_\Lambda,\hat\rho_\Lambda)$, which are defined in Eqs.~\eqref{eq:trunc}, \eqref{eq:linflow} and \eqref{eq:gradients}, are kept generic, and the class-E completion defined below fixes the frame in which the improvement scale is defined. In addition, the estimates that use class-J coefficients in Secs.~\ref{sec:matching} and \ref{sec:poles} are conditional examples and not predictions of a specific completion.

\subsection{Two classes of completions}
\label{sec:classes}

There are two classes of weakly nonlocal completions of the Starobinsky model in the literature, which we call class~J and class~E after the frame in which the completion is formulated. We distinguish them because not every ingredient used below is available in both. In class~J, the action is written in the Jordan frame with entire form factors acting on the curvatures, schematically $R\,F_R(\Boxop)R+C\,F_C(\Boxop)C$ \cite{Briscese2013,Modesto2013,Koshelev2016,KoshelevKumarStarobinsky2023}, where $C$ stands for the Weyl tensor and $F_R$, $F_C$ are the form factors. Here the scalaron is the scalar mode of the metric, and the tree-level tensor spectrum is dressed by the Weyl form factor. Moreover, for asymptotically polynomial form factors, the one-loop divergences are those of the corresponding polynomial theory \cite{ModestoRachwalShapiro2018}. In class~E, the completion is built on the Einstein-frame scalar-tensor action \cite{ModestoOrlando2026}. Only two properties of the completion are needed for the inflationary analysis, i.e., it preserves every solution of the local equations of motion, and it dresses the Hessian without introducing new zeros in the tree-level propagator.

The completion of Ref.~\cite{ModestoOrlando2026} adds to a local action a term quadratic in its own equations of motion. It reads
\begin{equation}
S_{\rm NL}=S_{\rm loc}
+\frac12\int\dd^4x\sqrt{-g}\;
E_i\left(e^{\Hf(\hD/\Lstar^2)}-1\right)^{i}{}_{k}
\left(\hD^{-1}\right)^{kj}E_j ,
\qquad
E_i=\frac{\delta S_{\rm loc}}{\delta\Phi^i},
\quad
\hD_{ij}=\frac{\delta E_i}{\delta\Phi^j},
\label{eq:SNL}
\end{equation}
where $\Phi^{i}$ collectively denotes the fields of the local action $S_{\rm loc}$, $\Lstar$ is the nonlocality scale, and $\Hf$ is an entire function with $\Hf(0)=0$, so that $e^{\Hf}$ has no zeros. There are two consequences which matter here. Firstly, the first variation is $\delta S_{\rm NL}=\delta\Phi\,e^{\Hf}E+\ord(E^{2})$. Therefore, any configuration with $E_i=0$ remains a solution of the completed theory, and the Starobinsky background is inherited exactly. Secondly, the quadratic form on shell is dressed as
\begin{equation}
\delta^2S_{\rm NL}\big|_{E=0}
=\delta\Phi\;\hD\,e^{\Hf(\hD/\Lstar^2)}\;\delta\Phi ,
\label{eq:hessdress}
\end{equation}
and, since $e^{\Hf}$ is zero free, the tree-level poles remain those of the local theory, namely $k^{2}=0$ and $k^{2}=-\Ms^{2}$, where $k^{2}\equiv k_\mu k^{\mu}$ is the momentum squared. The linearized perturbations obey the local equations, and hence the tree-level scalar and tensor spectra in class~E are those of the local model \cite{ModestoOrlando2026}. Note that the absence of additional zeros of the Hessian does not by itself fix the normalization of the two-point functions. That the normalization is also unchanged requires $\Hf(0)=0$ on shell. We rely on Ref.~\cite{ModestoOrlando2026} for this statement and have not rederived the quadratic actions of the perturbations. In this work, we take class~E as the working framework, because the improvement scale of Sec.~\ref{sec:rg} is then defined in the frame in which the completion is defined. The ingredients imported from class~J are identified where they enter.

In addition, two normalizations are required in Sec.~\ref{sec:poles}. We consider the Jordan-frame action $\tfrac12\mpl^{2}R+\alpha_CC^{2}+\alpha_RR^{2}$ and a perturbation $h_{\mu\nu}$ around flat space. In the transverse-traceless gauge, the linearized curvatures are $R_{\mu\nu}^{\rm lin}=-\frac12\Boxop h_{\mu\nu}$ and $R^{\rm lin}=0$. By using Eq.~\eqref{eq:C2identity}, one obtains
\begin{equation}
\Pi_2^{-1}\propto k^{2}\left[1-\frac{4\alpha_C}{\mpl^{2}}k^{2}\right],
\qquad
\Pi_0^{-1}\propto k^{2}\left[1+\frac{12\alpha_R}{\mpl^{2}}k^{2}\right],
\qquad
\alpha_R^{\rm tree}=\frac{\mpl^{2}}{12\Ms^{2}} ,
\label{eq:spin2norm}
\end{equation}
where $\Pi_2$ and $\Pi_0$ are the spin-2 and spin-0 parts of the propagator, and the spin-0 normalization follows from the $f(R)$ dictionary and the existence of the pole at $k^{2}=-\Ms^{2}$. Since the classical theory is unaltered by the completion, any deviation in the observables must originate in quantum corrections.

\subsection{One-loop running of the local sector and the improved potential}
\label{sec:rg}

In this subsection, we first isolate the local one-loop coefficients that feed the scalaron potential, and then place them on the slow-roll background through a scale prescription.

For a Laplace-type operator $\mathcal{D}=-\Boxop+\mathcal{E}$, the divergent part of the one-loop effective action is given by $\Gamma^{(1)}_{\rm div}=[2(4\pi)^2\bar\epsd]^{-1}\int\dd^4x\sqrt{-g}\,{\rm tr}\,a_2$. Here we use the standard Seeley--DeWitt coefficient \cite{BirrellDavies1982,BuchbinderBook,Vassilevich2003}, where the relevant contribution is the conformal square $\tfrac12(\mathcal{E}-R/6)^{2}$. Substituting $\mathcal{E}=V''+\xi R$ for the scalaron, we obtain
\begin{equation}
\Gamma^{(1)}_{{\rm div},\phi}
=\frac{1}{2(4\pi)^2\bar\epsd}\int\dd^4x\sqrt{-g}
\left\{\frac12\left[V''\right]^{2}
+\left(\xi-\frac16\right)R\,V''
+\frac12\left(\xi-\frac16\right)^{2}R^{2}+\cdots\right\},
\label{eq:scalarDiv}
\end{equation}
so that $\xi=1/6$ is a fixed point. Our conventions and the mapping from poles to beta functions are recorded in Appendix~\ref{app:hk}, namely $\mu\,\dd\lambda_O/\dd\mu=-X$ for $\Gamma_{\rm div}=+X/(2\bar\epsd)\int O$ with $\Gamma\supset+\int\lambda_O O$. This rule has been verified against the Coleman--Weinberg potential. Here $O$ labels a local operator in the truncation and $\lambda_O$ is its coefficient in the effective action. The pair $(\lambda_O,O)$ ranges over
\begin{equation}
\left(\tfrac12\mpl^{2},R\right),\quad
\left(\alpha_C,C^{2}\right),\quad
\left(\alpha_R,R^{2}\right),\quad
\left(\alpha_\GB,\GB\right),\quad
\left(-\rho_\Lambda,1\right),\quad
\left(-V,1\right),\quad
\left(-\tfrac12\xi,R\phi^{2}\right).
\label{eq:opbasis}
\end{equation}
The rule as stated applies when the operator enters $\Gamma$ with a positive coefficient. The final three entries of Eq.~\eqref{eq:opbasis} carry an explicit minus sign and therefore reverse it, which is the origin of $\be_V=+X_V$.

In addition, Eq.~\eqref{eq:scalarDiv} fixes how a matter sector feeds the running of the $R^{2}$ coefficient, and this effect will be used in Sec.~\ref{sec:matter}. Only the final term contributes, and it vanishes at $\xi=1/6$. Massless Dirac fermions and gauge vectors are conformal, and hence their $a_2$ coefficient carries no $R^{2}$ structure \cite{BirrellDavies1982,BuchbinderBook,Vassilevich2003}. Masses generate $m^{2}R$ and $m^{4}$ terms but no $R^{2}$ term. Therefore, the one-loop $R^{2}$ divergence in any matter sector originates from nonminimally coupled scalars alone. By using the rule above, their contribution to $\be_{\alpha_R}$ is negative and is given by
\begin{equation}
B_R^{\rm matter}=-\frac{1}{2(4\pi)^{2}}
\sum_s\left(\xi_s-\frac16\right)^{2} ,
\label{eq:BRmatter}
\end{equation}
where the sum runs over real scalar components with nonminimal couplings $\xi_s$. This means that the coefficient $\alpha_R$ grows toward the IR. Its role in models with a large Higgs nonminimal coupling is discussed in Refs.~\cite{SalvioMazumdar2015,GorbunovTokareva2019}. It was established in Ref.~\cite{Calcagni2023} that the nonlocal theory with nonminimally coupled matter is renormalizable. However, this does not by itself show that the heat kernel of a local spectator scalar equals the scalar contribution to the beta function of the full nonlocal scalar--gravity system. We therefore use Eq.~\eqref{eq:BRmatter} as an estimate of that contribution.

The local operators renormalized at one loop in the gravitational sector are listed in Ref.~\cite{ModestoOrlando2026}. In the Einstein-frame description, we follow the gravitational couplings $\{\mpl^{2},$ $\rho_\Lambda,$ $\alpha_C,$ $\alpha_R\}$ together with the scalaron parameters $\Ms^{2}$ and $\xi$, and we retain $\alpha_\GB$ for completeness. The beta functions take the forms
\begin{align}
\be_V&=\frac{[V'']^{2}}{2(4\pi)^{2}},
\qquad
\be_\xi=\Big(\xi-\tfrac16\Big)\frac{V''''}{(4\pi)^{2}},
\qquad
\be_{\Ms^{2}}=\gamma_M\Ms^{2},
\nonumber\\[4pt]
b_G&\equiv\frac{\dd\ln\mpl^{2}}{\dd t}=\frac{2B_G}{\mpl^{2}},
\qquad
\be_{\rho_\Lambda}=B_\Lambda,
\qquad
\be_{\alpha_{C,R}}=B_{C,R}.
\label{eq:trunc}
\end{align}
Here $\be_X\equiv\dd X/\dd t$ denotes the beta function of the coupling $X$, $\gamma_M$ and $b_G$ are the logarithmic running rates of $\Ms^{2}$ and $\mpl^{2}$, respectively, and $B_G$, $B_\Lambda$, $B_C$ and $B_R$ are constants. In addition, $\rho_\Lambda$ is the vacuum energy density and $\alpha_\GB$ is the coefficient of the Gauss--Bonnet term. Moreover, the scalar wavefunction renormalization vanishes at this order in the scalar loop on a fixed background. The scalar--gravity mixing and finite kinetic corrections in the completed theory are not included.

It is worth noting that this truncation does not close, and Eq.~\eqref{eq:trunc} is a projection. With the nonpolynomial potential in Eq.~\eqref{eq:Vst}, the divergence $\tfrac12[V'']^{2}=\tfrac12\Ms^{4}(2u^{2}-u)^{2}$ is a function of the field that is not proportional to $\Vst$, and it does not renormalize a single mass parameter. Likewise, $(\xi-\tfrac16)RV''$ is not proportional to $R\phi^{2}$. It generates a field-dependent coefficient of $R$, and $\be_\xi$ evaluated at a field value is a projection of it. The parts of the divergences that fall outside the six couplings are dropped, and their size sets the error of the projection. At the pivot scale, they are given by
\begin{equation}
\frac{\be_V}{\Vst}=\frac{2\Ms^{2}}{3(4\pi)^{2}\mpl^{2}}\,\frac{u^{2}(1-2u)^{2}}{(1-u)^{2}}\simeq1\times10^{-16},
\qquad
\frac{2\,|\xi-\tfrac16|\,|V''|}{(4\pi)^{2}\mpl^{2}}\simeq4\times10^{-15}\quad(\xi=0),
\label{eq:projection}
\end{equation}
per unit of the RG time $t$, for the potential and for the relative variation of the coefficient of $R$, respectively. Both are far below the gravitational parameters retained, and therefore the projection is adequate at the accuracy of the estimates below. However, this is not a statement that the six couplings give a complete local description. Moreover, the scale $\Lstar$ carries no beta function in minimal subtraction because divergences are local, whereas $\Lstar$ parametrizes an entire function. One can see from Eq.~\eqref{eq:betaxi} that $\be_\xi$ is numerically irrelevant, and we set $\xi=0$.

For class~J with a form factor of asymptotically polynomial behavior of degree $\Npol$, the gravitational coefficients are those of the polynomial theory $\sum_{n\le\Npol}(\omega_{n,R}R\Boxop^{n}R+\omega_{n,C}C\Boxop^{n}C)$ \cite{ModestoRachwalShapiro2018}. They read
\begin{align}
B_G&=\frac{1}{6(4\pi)^{2}}
\left(\frac{5\,\omega_{\Npol-1,C}}{\omega_{\Npol,C}}
+\frac{\omega_{\Npol-1,R}}{\omega_{\Npol,R}}\right),
\nonumber\\[4pt]
B_{\omega_{\rm cc}}&=\frac{1}{(4\pi)^{2}}
\left[\frac{5\,\omega_{\Npol-2,C}}{\omega_{\Npol,C}}
+\frac{\omega_{\Npol-2,R}}{\omega_{\Npol,R}}
-\frac52\frac{\omega_{\Npol-1,C}^{2}}{\omega_{\Npol,C}^{2}}
-\frac12\frac{\omega_{\Npol-1,R}^{2}}{\omega_{\Npol,R}^{2}}\right],
\label{eq:BGgen}
\end{align}
which are valid for $\Npol\ge2$ with vanishing generalized Gauss--Bonnet couplings, and one-loop exact for $\Npol\ge3$. The cases $\Npol=0$ and $\Npol=1$ are not obtained as limits of Eq.~\eqref{eq:BGgen} \cite{ModestoRachwalShapiro2018,RachwalModestoPinzulShapiro2021}. With the entry $(-\rho_\Lambda,1)$ of Eq.~\eqref{eq:opbasis}, one has $B_\Lambda=-B_{\omega_{\rm cc}}$, where $B_{\omega_{\rm cc}}$ is the beta function of the coefficient $\omega_{\rm cc}$ of the unit operator in that reference. The coefficients $\omega_{n,R}$, $\omega_{n,C}$ and the signs of Eq.~\eqref{eq:BGgen} are those of Ref.~\cite{ModestoRachwalShapiro2018}, translated to our conventions for the Einstein--Hilbert and vacuum terms, which are the conventions of Ref.~\cite{ModestoOrlando2026} (see Appendix~\ref{app:hk}). More importantly, the content of Eq.~\eqref{eq:BGgen} is clearest in terms of the zeros of the asymptotic polynomials. Let the polynomial of the spin-$j$ sector be proportional to $\prod_{i=1}^{\Npol}(k^{2}+m_{i,j}^{2})$, where $k^{2}$ is the Euclidean momentum squared, and $j=2$ and $j=0$ refer to the Weyl sector and the scalar sector, respectively. In each sector, one has $\omega_{\Npol-1}/\omega_{\Npol}=\sum_im_i^{2}$ and $\omega_{\Npol-2}/\omega_{\Npol}=\sum_{i<j}m_i^{2}m_j^{2}$, so that
\begin{equation}
B_G=\frac{1}{6(4\pi)^{2}}\bigg[5\sum_i m_{i,2}^{2}+\sum_i m_{i,0}^{2}\bigg],
\qquad
B_\Lambda=\frac{1}{2(4\pi)^{2}}\bigg[5\sum_i m_{i,2}^{4}+\sum_i m_{i,0}^{4}\bigg].
\label{eq:BGmass}
\end{equation}
This means that each zero contributes as one minimally coupled real scalar of mass $m_i$ (see Table~\ref{tab:signs}), with multiplicity five in the spin-two sector and one in the spin-zero sector. Note that the $m_i^{2}$ are parameters of the form factor and not masses of states, since $e^{\Hf}$ has no zeros. They are real or form complex-conjugate pairs. If they are real, then $B_\Lambda>0$, and if they are also positive, then $B_G>0$. For complex pairs, either sign occurs. For example, the equal ratios $\omega_{\Npol-1}/\omega_{\Npol}=\Lstar^{2}$ and $\omega_{\Npol-2}/\omega_{\Npol}=\Lstar^{4}$ give $\sum_im_i^{4}=-\Lstar^{4}$ in each sector and $B_\Lambda<0$. In addition, if the asymptotic polynomial is a monomial, all zeros lie at the origin and $B_G=B_\Lambda=0$. For the class-E completion in Eq.~\eqref{eq:SNL}, whose UV operator is built from the Hessian of the Einstein-scalar action, the corresponding coefficients have not been computed to our knowledge. We use Eq.~\eqref{eq:BGgen} for its dimensional structure, which is all that the estimates of Sec.~\ref{sec:matching} require, and we regard the sign pattern of Eq.~\eqref{eq:BGmass} as indicative for that class. Note also that the calculation of Ref.~\cite{ModestoRachwalShapiro2018} is not the mixed scalar--metric Hessian of the Einstein-frame completion. We refer the readers to that reference for the physical interpretation of this minimal-subtraction running and the identification of the scale that it requires. For constant $B_i$, the flow integrates to
\begin{equation}
\begin{aligned}
\mpl^{2}(\mu)&=\mpl^{2}\left[1+b_G t\right],
&\rho_\Lambda(\mu)&=\rho_\Lambda+B_\Lambda t ,\\
\alpha_R(\mu)&{}=\alpha_R+B_R t ,
&\Ms^{2}(\mu)&=\Ms^{2}\left[1+\gamma_M t{}+\ord(t^{2})\right],
\end{aligned}
\label{eq:linflow}
\end{equation}
where the first three relations are exact in $t$ when the beta functions are one-loop exact. The last one is of first order, because $\Ms^{2}$ is related to the ratio $\mpl^{2}/\alpha_R$ in Sec.~\ref{sec:matter}. The logarithmic derivative $\gamma_M(t)=b_G/(1+b_Gt)-(B_R/\alpha_R)/[1+(B_R/\alpha_R)t]$ is then not constant, and $\gamma_M$ below denotes its value at $t=0$. These four running parameters are the only inputs required for our results.

\begin{figure}[t]
\centering
\includegraphics[width=0.95\textwidth]{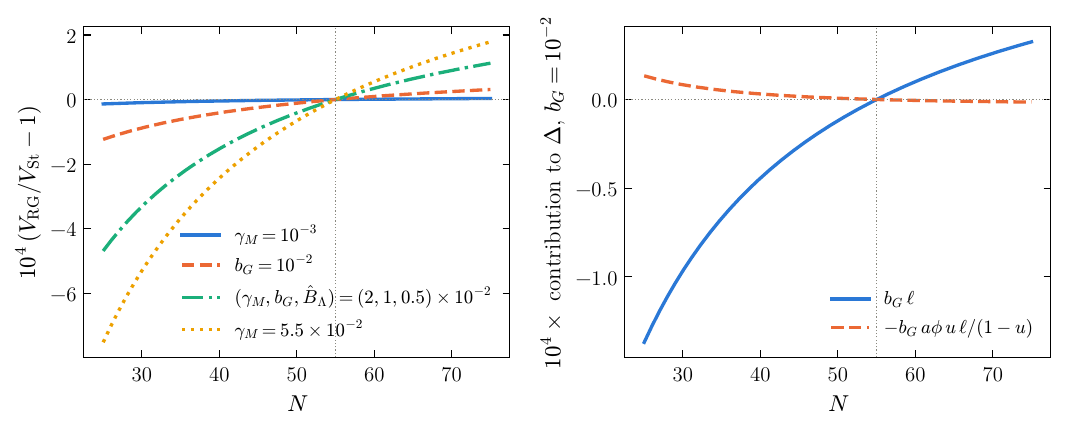}
\caption{Left: $\VRG/\Vst-1$ across the observable window, evaluated along the tree-level trajectory, the deformation vanishing at the pivot scale $N_*=55$ by construction. The parameter values are illustrations. Those of $\gamma_M$ and $b_G$ exceed the matched values of Sec.~\ref{sec:matching} by at least four orders of magnitude and are not restricted to $\gamma_M=b_G$. Right: the $b_G$ channel split into $b_G\ell$ and the exponent shift.}
\label{fig:pot}
\end{figure}

The improvement scale determines the functional form of the correction, and we state our choice as an assumption. Resumming leading logarithms into an inflaton potential is a standard construction \cite{PozdeevaVernov2016}, and implementations differ in the choice of argument. For the gravitational sector on a slow-roll background, we adopt the Einstein-frame Hubble rate, i.e., $\mu=H(\phi)$ with $H^{2}=\Vst/3\mpl^{2}$, anchored at $\mu_0=H(\phi_0)$ at the pivot scale. This is the curvature scale of the frame in which the class-E completion is defined, and $\Lstar$ is constant in that frame. We will examine in Sec.~\ref{sec:scheme} the scales defined in other ways, including the Jordan-frame curvature, for which the conclusions change. With this choice, one obtains
\begin{equation}
\ell(\phi)\equiv\ln\frac{\mu(\phi)}{\mu_0}
=\ln\frac{1-u}{1-u_0}
=\frac12\ln\frac{\Vst(\phi)}{\Vst(\phi_0)} ,
\label{eq:ell}
\end{equation}
where the final equality is exact. The results of Sec.~\ref{sec:results} rest on this identity. The logarithm is small throughout the observable window, i.e., $|\ell|<0.005$ for $40\le N\le75$, and hence the construction does not resum large logarithms. It is a first-order parametric improvement, in which the running parameters are substituted into the tree-level potential, and we treat this substitution as an ansatz. A calculation of an observable combines the implicit running with the explicit $\mu$ dependence of the loop terms, schematically $(\mu\partial_\mu+\be_i\partial_{g_i}-\gamma_\phi\phi\partial_\phi)\Gamma=0$, where $g_i$ denotes the couplings and $\gamma_\phi$ is the anomalous dimension of the field. Here we retain the implicit running and drop the finite terms without a bound on their effect. Because $\ell$ vanishes at the pivot scale and is small across the observable interval, there is no logarithmic enhancement that would make the retained terms dominate the omitted finite ones. Substituting Eq.~\eqref{eq:linflow} into Eq.~\eqref{eq:Vst} and adding the vacuum energy, we obtain
\begin{equation}
\VRG=\Vst\left[1+\Delta\right],
\qquad
\Delta=\left(\gamma_M+b_G\right)\ell
-b_G\,\frac{a\phi\,u}{1-u}\,\ell
+\frac{\rho_\Lambda+B_\Lambda\ell}{\Vst}
+\frac{\Delta V_{\rm 1loop}}{\Vst} ,
\label{eq:Delta}
\end{equation}
where the second term arises from the running of $\mpl$ inside the exponential. This substitution lets $\mpl$ run in the potential while the gravitational coupling of the Einstein frame and the kinetic term are held fixed. The consistent frame normalization will be examined in Sec.~\ref{sec:framenorm}. The Coleman--Weinberg contribution $\Delta V_{\rm 1loop}$ is $\ord(10^{-16})$ for $\xi=0$ and $\ord(10^{-12})$ even for $\xi=1$. We therefore neglect it, in agreement with the corresponding conclusion for the local model \cite{Ellis2025}. Upon differentiation, all four running parameters enter the dynamics at leading order through a single combination, viz.,
\begin{equation}
\Delta'=\frac{a\,u}{1-u}\,K+\ord(u^{2}\ln u),
\qquad
K\equiv\gamma_M+b_G+\hat B_\Lambda-2\hat\rho_\Lambda,
\qquad
\frac{\Vst}{\Vst'}\Delta'=\frac{K}{2}{}+\ord(u\ln u),
\label{eq:gradients}
\end{equation}
where $\hat X\equiv X/V_0$. Interestingly, the vacuum-energy terms contribute to $K$ on the same footing as the two mass parameters. In Fig.~\ref{fig:pot}, we show the resulting deformation across the observable window for illustrative values of the running parameters, together with the split of the $b_G$ channel into its leading part and the exponent shift.

\subsection{Frame normalization of the running Einstein--Hilbert coefficient}
\label{sec:framenorm}

If the scale is promoted to $\mu(\phi)$ at the level of the action, the coefficient of $R$ becomes field dependent. The potential slow-roll formulae then apply only after a transformation to a frame with constant Planck mass, and the field must be canonically normalized there. The improved local action takes the form
\begin{equation}
\Gamma=\int\dd^4x\sqrt{-g}\left[\frac12F(\phi)R-\frac12Z(\phi)(\nabla\phi)^{2}-U(\phi)\right],
\qquad
F=\mpl^{2}\left[1+b_G\ell\right],\quad Z=1,
\label{eq:improvedaction}
\end{equation}
with $U=V_0\left[1+(\gamma_M+b_G)\ell\right]\left(1-e^{-a_\mu\phi}\right)^{2}+\rho_\Lambda+B_\Lambda\ell$, since $V_0=\tfrac34\mpl^{2}\Ms^{2}$. For the exponent, we write $a_\mu=a\,[1-\tfrac12\sigma b_G\ell]$. Here the value $\sigma=1$ applies if the exponent keeps its tree-level relation $a_\mu=\sqrt{2/3}\,F^{-1/2}$ to the coefficient of $R$, while $\sigma=0$ applies if it is an independent coupling of the Einstein-frame action without a beta function in the truncation. In this work, we adopt $\sigma=1$, and there are three reasons for this choice. Firstly, the exponent is defined through the Planck mass in Eq.~\eqref{eq:Vst}, and the truncation of Eq.~\eqref{eq:trunc} contains no independent coupling for it. Secondly, the relation $a=\sqrt{2/3}/\mpl$ belongs to the $f(R)$ dictionary, which we also assume for the scalaron mass in Sec.~\ref{sec:matter}. Thirdly, with $\sigma=1$ the improved potential remains the Einstein-frame image of $R+R^{2}$ gravity, whereas $\sigma=0$ moves the model to an $\alpha$-attractor E-model with $\alpha\neq1$. We keep $\sigma$ in the following expressions in order to quantify the dependence on this choice. By using the rescaling $\tilde g_{\mu\nu}=(F/\mpl^{2})g_{\mu\nu}$, the action can be rewritten as
\begin{equation}
\Gamma=\int\dd^4x\sqrt{-\tilde g}\left[\frac12\mpl^{2}\tilde R-\frac12\mathcal{K}(\phi)(\tilde\nabla\phi)^{2}-V_E(\phi)\right],
\quad
\mathcal{K}=\frac{\mpl^{2}Z}{F}+\frac{3\mpl^{2}}{2}\frac{F'^{2}}{F^{2}},
\quad
V_E=\frac{\mpl^{4}U}{F^{2}} .
\label{eq:einsteinframe}
\end{equation}
At first order in the running parameters, one finds $\mathcal{K}=1-b_G\ell$ and $V_E=\Vst[1+\Delta_E]$ with
\begin{equation}
\Delta_E=\left(\gamma_M-b_G\right)\ell-\sigma\,b_G\,\frac{a\phi\,u}{1-u}\,\ell+\frac{\rho_\Lambda+B_\Lambda\ell}{\Vst},
\qquad
K_E\equiv\gamma_M-b_G+\hat B_\Lambda-2\hat\rho_\Lambda .
\label{eq:DeltaE}
\end{equation}
There are two differences from Eq.~\eqref{eq:Delta}. Firstly, the running of $\mpl^{2}$ enters the removable combination with the opposite sign. The reason is that the plateau height in Planck units is $V_0/\mpl^{4}=1/16\alpha_R$. A running of $\mpl^{2}$ alone is a change of units and, by using Eq.~\eqref{eq:gammaMfix} below, the mass parameters contribute to $K_E$ only through $\gamma_M-b_G=-B_R/\alpha_R=\ord(10^{-11})$. Secondly, a kinetic factor $\mathcal{K}=1+\delta_{\mathcal K}$ with $\delta_{\mathcal K}=-b_G\ell$ appears. At fixed field, it gives $\eps=\eps_0(1+2\Rc-\delta_{\mathcal K})$ and $\dd N=\dd N_0(1-\Rc+\delta_{\mathcal K})$, and Eq.~\eqref{eq:response} generalizes as stated in Eq.~\eqref{eq:app_kinetic}. Since $\delta_{\mathcal K}=\ord(u)$, the kinetic factor is of class~S. Numerically, at $N_*=55$, we obtain
\begin{equation}
\Delta n_s=-5.15\times10^{-4}\left(\gamma_M-b_G\right)-c_{\mathcal K}\,b_G+\cdots,
\qquad
c_{\mathcal K}=
\begin{cases}
9.97\times10^{-4} & (\sigma=0),\\
16.96\times10^{-4} & (\sigma=1),
\end{cases}
\label{eq:frameresult}
\end{equation}
where the dots stand for the vacuum-energy channels, which are unchanged at this order because the additional terms $-2b_G(\hat\rho_\Lambda+\hat B_\Lambda\ell)\ell$ are of second order. For the adopted value $\sigma=1$, the response to $b_G$ at fixed $\gamma_M$ is therefore $-11.8\times10^{-4}$, against $-12.1\times10^{-4}$ from the substitution in Eq.~\eqref{eq:Delta}. In the direction $\gamma_M=b_G$, the two values are $-17.0\times10^{-4}$ and $-17.3\times10^{-4}$, respectively. For comparison, $\sigma=0$ would give $-4.8\times10^{-4}$ and $-10.0\times10^{-4}$.

According to these results, the substitution in Eq.~\eqref{eq:Delta} reproduces the frame-consistent response to $b_G$ within $3\%$, although it attributes the effect to a running prefactor and not to the kinetic normalization. If the exponent did not run, the response would be smaller by a factor of about $2.5$. In both cases, the $\mpl$ channel belongs to classes R and S, its contribution to $n_s$ is $\ord(b_G\ln N/N^{2})$, and for the matched values $b_G\le6\times10^{-7}$ of Sec.~\ref{sec:matching} it is of order $10^{-9}$. More importantly, the classification, the vacuum-energy channels and the estimates of Sec.~\ref{sec:matching} do not depend on this choice. The tables and figures below use the substitution in Eq.~\eqref{eq:Delta}, whose $b_G$ entries agree with the adopted choice $\sigma=1$ within $3\%$.

Note that we have not derived $F$, $Z$ and $U$ from the one-loop effective action of a definite completion. In particular, divergent or finite contributions to $Z$ are not included, and the frame dependence of the off-shell one-loop divergences \cite{KamenshchikSteinwachs2015} is not addressed.

\section{Application to the gravitational flow}
\label{sec:results}

\subsection{Trajectory of the flow and the role of one-loop exactness}
\label{sec:pfamily}

According to Eq.~\eqref{eq:gradients}, the improvement multiplies $\Vst$ by $1+K\ell$ at first order in the running parameters. Here the combination $K$ is defined by this expansion. Note that a flow equation $\dd V/\dd t=KV$ with constant $K$ does not follow from the system of independent couplings, which contains the additive vacuum term and the exponent deformation. By using the identity in Eq.~\eqref{eq:ell}, this factor is turned into a power of the tree-level potential, i.e., $1+K\ell=(\Vst/\Vst(\phi_0))^{K/2}+\ord(K^{2}\ell^{2})$, so that
\begin{equation}
\VRG\;\propto\;\left(1-e^{-a\phi}\right)^{p}\left[1+\ord(u\ln u)\right]\left[1+\ord(K^{2}\ell^{2})\right],
\qquad
p=2+K .
\label{eq:power}
\end{equation}
At first order in $K$, the flow therefore moves the model along the E-model family $V_p=A(1-e^{-a\phi})^{p}$. For that family, the slow-roll parameters, the number of $e$-folds and the end point read
\begin{equation}
\eps_V=\frac{p^{2}u^{2}}{3(1-u)^{2}},
\qquad
\eta_V=\frac{2p\,u\,(pu-1)}{3(1-u)^{2}},
\qquad
N=\frac{3}{2p}\left(\frac1u+\ln u\right)+{\rm const},
\qquad
u_{\rm end}=\frac{\sqrt3}{p+\sqrt3},
\label{eq:pexact}
\end{equation}
from which we obtain $u_N=3/(2pN)$. Interestingly, every factor of $p$ then cancels between the slow-roll parameters and the mapping to the number of $e$-folds, and one finds
\begin{equation}
\eps_V\to\frac{3}{4N^{2}},
\qquad
\eta_V\to-\frac1N,
\qquad
n_s\to1-\frac2N,
\qquad
r\to\frac{12}{N^{2}} ,
\label{eq:universal}
\end{equation}
which reproduces the exponent independence established in Refs.~\cite{KalloshLindeRoest2013,KalloshLindeRoest2014,Roest2014}. Here these observables are evaluated at fixed $N$ derived from the deformed potential, and hence the shift in the field-to-$e$-fold mapping is included. The $p$-family does not belong to the exactly equivalent class of Eq.~\eqref{eq:exactclass}. A field shift maps $(1-u)^{2}$ to $(1-\lambda u)^{2}$, which reproduces $(1-u)^{p}$ at order $u$ for $\lambda=p/2$ and fails at order $u^{2}$ unless $p=2$. The deformation is therefore invisible at leading order and visible beyond it, and this makes the residual computation of Sec.~\ref{sec:analytic} necessary.

We now explain how the one-loop exactness enters. For $\gamma_{\rm uv}>D/2$, where $e^{\Hf(z)}\sim z^{\gamma_{\rm uv}+1}$ at large $z$ and $z$ denotes the dimensionless argument of the form factor, which is equivalent to polynomial theories with $\Npol\ge3$, the superficial degree of divergence $\delta(G)=D-2\gamma_{\rm uv}(L-1)$ is negative for every graph $G$ with $L\ge2$ loops. This means that the local UV divergences occur at one loop alone \cite{ModestoRachwal2014,ModestoRachwalShapiro2018,ModestoOrlando2026}. The beta functions of Eq.~\eqref{eq:trunc} then give the local divergent running within the projected truncation of Sec.~\ref{sec:rg}, and $\mpl^{2}$, $\rho_\Lambda$ and $\alpha_R$ are exactly linear in $t$. However, linear running does not exponentiate, and Eq.~\eqref{eq:power} is a first-order statement. The termination of the UV divergences at one loop does not remove the nonlinear dependence on the couplings or the finite loop terms. The coefficient of the second-order term depends on the matching and renormalization conditions imposed at that order, which we have not specified, and we quote it for the linear running in Eq.~\eqref{eq:linflow} only. By using $\Ms^{2}=\mpl^{2}/12\alpha_R$, which anticipates Sec.~\ref{sec:matter}, the amplitude is given by
\begin{equation}
\frac34\mpl^{2}(\mu)\Ms^{2}(\mu)=\frac{\mpl^{4}(\mu)}{16\,\alpha_R(\mu)}
=V_0\,\frac{(1+b_G\ell)^{2}}{1+(B_R/\alpha_R)\,\ell}
=V_0\left[1+\left(b_G+\gamma_M\right)\ell+\gamma_M^{2}\,\ell^{2}+\ord(\ell^{3})\right],
\label{eq:product}
\end{equation}
with $\gamma_M=b_G-B_R/\alpha_R$. If instead $\mpl^{2}$ and $\Ms^{2}$ are both taken linear, the coefficient of $\ell^{2}$ is $b_G\gamma_M$, and the two expressions coincide for $\gamma_M=b_G$. The quadratic term belongs to class~S, since $\Rc[c\,\ell^{2}]=c\,\ell=\ord(u)$. For the illustrative coefficient $c=6\times10^{-4}$, corresponding to $b_G=2\times10^{-2}$ and $\gamma_M=3\times10^{-2}$, it contributes $\Delta n_s=-5.2\times10^{-7}$ at $N_*=55$ against $-2.42\times10^{-5}$ from the linear term with $K=5\times10^{-2}$, i.e., a relative correction of about two percent. For the matched values of Sec.~\ref{sec:matching}, it is smaller by many orders of magnitude. Finite nonlocal contributions remain outside this truncation.

\subsection{Matter content and the running of the scalaron mass}
\label{sec:matter}

The couplings of Eq.~\eqref{eq:trunc} are not independent if the Einstein-frame flow of the scalaron mass is the image of the Jordan-frame flow of the $R^{2}$ coefficient. In class~J, the scalaron is the scalar mode of the $R^{2}$ operator, and Eq.~\eqref{eq:spin2norm} gives $\Ms^{2}=\mpl^{2}/12\alpha_R$. In addition, the one-loop divergences computed in the two frames agree on shell \cite{KamenshchikSteinwachs2015}. We adopt this relation as an assumption for class~E, where $\alpha_R$ then denotes the Jordan-frame coefficient and $\delta\alpha_R$ of Sec.~\ref{sec:poles} is an additional Einstein-frame operator. It implies
\begin{equation}
\gamma_M=\frac{\dd\ln\Ms^{2}}{\dd t}
=b_G-\frac{B_R}{\alpha_R} .
\label{eq:gammaMfix}
\end{equation}
The first term is gravitational and already contributes to $K$, while the second term carries the matter dependence and has a tightly constrained denominator. By using Eq.~\eqref{eq:omega0Rvalue}, the measured scalar amplitude fixes $\alpha_R=\omega_{0,R}=5.6\times10^{8}$. This large value renders the matter channel numerically negligible. Having used Eq.~\eqref{eq:BRmatter}, a Standard Model (SM) scalar sector of one complex Higgs doublet with four real components sharing a nonminimal coupling $\xi_H$ leads to
\begin{equation}
-\frac{B_R^{\rm matter}}{\omega_{0,R}}
=\frac{4\left(\xi_H-\tfrac16\right)^{2}}{2(4\pi)^{2}\omega_{0,R}}
=2.3\times10^{-11}\left(\xi_H-\frac16\right)^{2} .
\label{eq:SMgamma}
\end{equation}
The gravitational part of $B_R$ is a pure number of order $(4\pi)^{-2}$. In the six-derivative theory, it equals $-7/36(4\pi)^{2}$ \cite{RachwalModestoPinzulShapiro2021}, which gives $|B_R|/\omega_{0,R}=2\times10^{-12}$. Both contributions are therefore $\ord(10^{-11})$ or smaller for $\xi_H=\ord(1)$, and we obtain
\begin{equation}
\gamma_M=b_G+\ord(10^{-11}),
\qquad
K=2b_G+\hat B_\Lambda-2\hat\rho_\Lambda+\ord(10^{-11}) .
\label{eq:Kfixed}
\end{equation}
Relative to $b_G$, the correction is $\ord(10^{-5})$ at the upper edge of the window in Eq.~\eqref{eq:window_L}, where $b_G=6\times10^{-7}$. Near the lower edge, one has $b_G=3\times10^{-11}$ and the two terms of Eq.~\eqref{eq:gammaMfix} are comparable, but both are then irrelevant for the observables. This means that the truncation possesses one fewer independent direction than Eq.~\eqref{eq:trunc} suggests. Tables~\ref{tab:validate} and \ref{tab:resid} are unaffected, since they are a channel-by-channel decomposition. The matter content fixes the difference $\gamma_M-b_G$. It does not fix $K$, which is dominated by the vacuum-energy terms. The matter contribution to the removable channel is given by
\begin{equation}
\Delta n_s^{\rm matter}
=-1.2\times10^{-14}\left(\xi_H-\frac16\right)^{2}
\qquad (N_*=55) ,
\label{eq:SMshift}
\end{equation}
which is obtained with the linear-response coefficient $-5.15\times10^{-4}$ of the $\gamma_M$ channel. Even at $\xi_H=10^{4}$, i.e., a value of the size used in Higgs inflation, its magnitude reaches only $1.2\times10^{-6}$, which is more than two orders of magnitude below the maximum shift of Eq.~\eqref{eq:maxshift} and nearly four orders below the displacement the ACT data require. Note that this extrapolation treats the Higgs field as a spectator. A large nonminimal coupling does not by itself determine which field drives inflation, and a two-field analysis would be needed in that regime. The sign of Eq.~\eqref{eq:SMshift} follows from the dictionary of Table~\ref{tab:signs}.

However, the SM matter is not thereby invisible. Matter enters the scalaron potential a second time through the Coleman--Weinberg terms with inflaton-dependent masses. Such terms are not removable and will be discussed in Sec.~\ref{sec:classification}. They provide the mechanism by which the model predictions are altered in Refs.~\cite{AttractorsRadACT,Ellis2025}. Our statement is narrower, i.e., the matter contribution to the running of $\Ms^{2}$ is fixed by $\xi_H$ alone and is negligible.

\subsection{Residual shifts and the \texorpdfstring{$e$}{e}-fold channel}
\label{sec:analytic}

The leading cancellation is exact only at the order at which the E-model exponent is invisible. In order to obtain a prediction, one must keep the subleading field-to-$N$ relation and then add the reheating contribution, which is not protected by the potential criterion.

Comparing $\VRG$ and $\Vst$ at a fixed field value, one obtains $\Delta n_s|_\phi=-\tfrac43Ku\simeq-K/N$. However, this is not observable, because the pivot scale is defined at a fixed number of $e$-folds. According to Eq.~\eqref{eq:gradients}, the deformation displaces the field-to-$e$-fold mapping by $\delta N=-\tfrac{K}{2}N$, which cancels that shift, in accordance with Eq.~\eqref{eq:average}. The physical residual therefore arises from the subleading term. Inverting the $N$-mapping as detailed in Appendix~\ref{app:inv} with $u_N=x(1+x\ln x+B_1x)$, $x=3/(2pN)$, and $B_1=\ln(p+\sqrt3)-p/\sqrt3-\tfrac12\ln3-1$, the large-$N$ expansion of the E-model predictions \cite{Roest2014,GarciaBellidoRoest2014,Linder2021} becomes explicitly dependent on $p$. It reads
\begin{equation}
n_s=1-\frac2N+\frac{1}{N^{2}}\left[\sqrt3-\frac32+g(p)\right]+\ord\left(\frac{\ln^{2}N}{N^{3}}\right),
\qquad
g(p)=\frac3p\left[\ln\frac{2Np}{\sqrt3(p+\sqrt3)}-1\right],
\label{eq:nsp}
\end{equation}
\begin{equation}
r=\frac{12}{N^{2}}-\frac{36}{pN^{3}}\left[\ln\frac{2pN}{\sqrt3(p+\sqrt3)}+\frac{p}{\sqrt3}\right]+\ord\left(\frac{\ln^{2}N}{N^{4}}\right),
\label{eq:rp}
\end{equation}
so that the shift induced by the parameter change $p=2\to2+K$ follows analytically as
\begin{equation}
\begin{aligned}
\Delta n_s^{\rm shape}&=\frac{g(2+K)-g(2)}{N^{2}}
\;\xrightarrow[K\ll1]{}\;
\frac{K}{N^{2}}\left[1.458-\frac34\ln N\right],\\
\left.\frac{\Delta r}{r}\right|^{\rm shape}
&{}=\frac{3K}{4N}\left[\ln\frac{4N}{3+2\sqrt3}+3-2\sqrt3\right]
=\frac{3K}{4N}\left[\ln N-0.944\right],
\end{aligned}
\label{eq:dnsanalytic}
\end{equation}
where the constant in $\Delta n_s^{\rm shape}$ is $\tfrac34[2\sqrt3-2+\ln((3+2\sqrt3)/4)]$. It is worth noting that both expressions also follow from Eq.~\eqref{eq:response} with $\Rc=K/2$, which provides an independent check, and the end-point term of that equation accounts for $0.464$ of the constant $1.458$. The shift of $n_s$ is $\ord(K\ln N/N^{2})$, i.e., one power of $1/N$ below the fixed-field estimate. Here Table~\ref{tab:validate} tests two statements. Firstly, Eq.~\eqref{eq:dnsanalytic} is the large-$N$ expansion of the E-model, and it agrees with the numerical E-model shift to $0.7\%$ at $N=55$ and $1.9\%$ at $N=45$ for both values of $K$. Secondly, the linearized potential $\Vst[1+K\ell]$ and the E-model differ at $\ord(K^{2})$, by $0.8\%$ at $K=0.01$ and $4.4\%$ at $K=0.05$ for $N=55$. The table therefore validates the large-$N$ expansion and the first-order correspondence, and not an exact resummed trajectory. In addition, the expression for $\Delta r/r$ agrees with the numerical solution to $2\%$ at $N_*=55$. The sign of $\Delta n_s^{\rm shape}$ remains negative for $N\gtrsim7$. This means that a positive $K$ lowers $n_s$ and drives the prediction away from the ACT preferred value.

\begin{table}[t]
\centering
\begin{tabular}{ccccccc}
\toprule
$N$ & $K$ & $\Vst[1+K\ell]$ & $V_p$, $p=2+K$ & Eq.~\eqref{eq:dnsanalytic} & ratio lin. & ratio $V_p$\\
\midrule
45 & 0.01 & $-6.946\times10^{-6}$ & $-7.004\times10^{-6}$ & $-6.871\times10^{-6}$ & 1.011 & 1.019\\
45 & 0.05 & $-3.302\times10^{-5}$ & $-3.445\times10^{-5}$ & $-3.381\times10^{-5}$ & 0.977 & 1.019\\
55 & 0.01 & $-5.085\times10^{-6}$ & $-5.128\times10^{-6}$ & $-5.094\times10^{-6}$ & 0.998 & 1.007\\
55 & 0.05 & $-2.416\times10^{-5}$ & $-2.522\times10^{-5}$ & $-2.506\times10^{-5}$ & 0.964 & 1.006\\
65 & 0.01 & $-3.907\times10^{-6}$ & $-3.941\times10^{-6}$ & $-3.943\times10^{-6}$ & 0.991 & 0.999\\
65 & 0.05 & $-1.856\times10^{-5}$ & $-1.938\times10^{-5}$ & $-1.939\times10^{-5}$ & 0.957 & 0.999\\
\bottomrule
\end{tabular}
\caption{Validation of Eq.~\eqref{eq:dnsanalytic} in the pure $\gamma_M$ channel, where $K=\gamma_M$. Columns three and four are the numerical shifts $\Delta n_s$ for the linearized potential $\Vst[1+K\ell]$ and for the E-model $V_p$. The analytic column uses $[g(2+K)-g(2)]/N^{2}$ without linearizing in $K$. The last two columns are the ratios of the numerical shifts to the analytic one.}
\label{tab:validate}
\end{table}

For the complete deformation in Eq.~\eqref{eq:Delta}, the shifts are linear combinations of four channel responses,
\begin{equation}
\Delta n_s=\sum_i\theta_i\,\frac{\nu_i(N)}{N^{2}},
\qquad
\frac{\Delta r}{r}=\sum_i\theta_i\,\frac{\varrho_i(N)}{N},
\qquad
\theta_i=(\gamma_M,b_G,\hat B_\Lambda,\hat\rho_\Lambda),
\label{eq:residual}
\end{equation}
and Eq.~\eqref{eq:response} reduces each $\nu_i$ and $\varrho_i$ to a one-dimensional quadrature. Their leading large-$N$ forms are given by
\begin{align}
\nu_{\gamma_M}&=1.458-\tfrac34\ln N,
&\nu_{\hat\rho_\Lambda}&=1.285-\tfrac32\ln N,
\nonumber\\
\nu_{\hat B_\Lambda}&=-0.835+\tfrac94\ln N,
&\nu_{b_G}&=-\tfrac34\ln^{2}N+2.57\ln N-1.62,
\label{eq:asymptotic}
\end{align}
where the coefficients of the logarithms are analytic, e.g., $\tfrac34(5-2\ln\tfrac43)-\tfrac34=2.57$ in the last case, and the constants of the last three channels were obtained by evaluating Eq.~\eqref{eq:response} up to $N=10^{8}$. All four channels shift $n_s$ at order $\ln N/N^{2}$, or $\ln^{2}N/N^{2}$ for $b_G$, in agreement with the classification of Sec.~\ref{sec:iff}. The $\gamma_M$ channel is the constant-$\Rc$ direction. The vacuum-energy channels have $\Rc=\hat B_\Lambda(\tfrac12-\ell)/(1-u)^{2}$ and $\Rc=-\hat\rho_\Lambda/(1-u)^{2}$, which are constant up to $\ord(u)$ and belong to class~R. Their $\ord(u)$ variation changes the coefficient of $\ln N$ from $-\tfrac34K$ to $+\tfrac94\hat B_\Lambda$ and $-\tfrac32\hat\rho_\Lambda$, respectively. Moreover, the exponent shift in the $b_G$ channel has $\Rc=\ord(u\ln u)$, belongs to class~S, and generates the $\ln^{2}N$ term. At $N_*=55$, the forms in Eq.~\eqref{eq:asymptotic} deviate from the numerical linear response by $0.7\%$, $3\%$, $5\%$ and $8\%$ for $\gamma_M$, $\hat\rho_\Lambda$, $\hat B_\Lambda$ and $b_G$, respectively. The slow convergence in the last channel reflects corrections of relative order $\ln N/N$ with large coefficients.

\begin{figure}[t]
\centering
\includegraphics[width=0.95\textwidth]{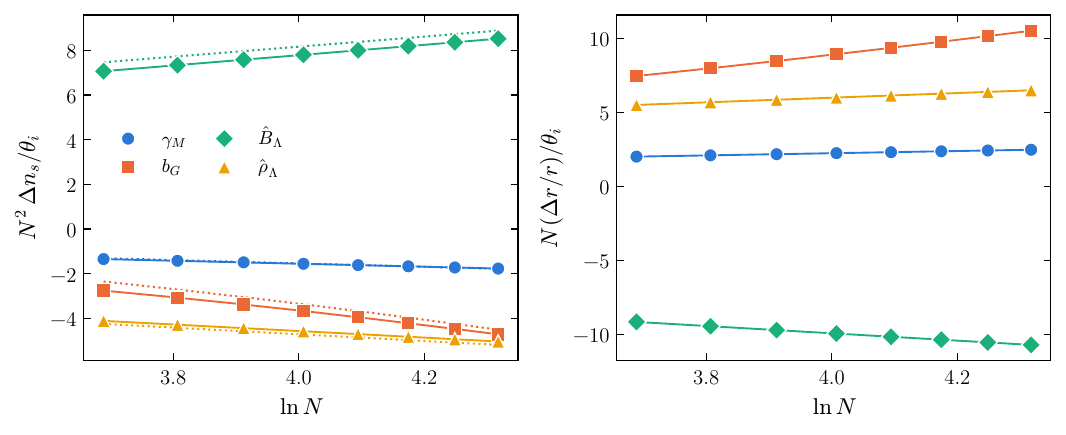}
\caption{Rescaled linear responses $N^{2}\Delta n_s/\theta_i$ (left) and
$N(\Delta r/r)/\theta_i$ (right) against $\ln N$. The points are the
numerical solution of Appendix~\ref{app:num} at small amplitude, which coincides with Eq.~\eqref{eq:response}. The solid lines are the
fits of Table~\ref{tab:resid}. The dotted lines in the left panel are the asymptotic forms \eqref{eq:asymptotic}. The $b_G$ points are not
collinear in $\ln N$, because that channel contains $\ln^{2}N$.}
\label{fig:resid}
\end{figure}

For numerical use over the observable window, we also provide the interpolating fits,
\begin{equation}
\nu_i(N)\simeq n^{(0)}_i+n^{(1)}_i\ln N+\frac{n^{(2)}_i}{N},
\qquad
\varrho_i(N)\simeq r^{(0)}_i+r^{(1)}_i\ln N+\frac{r^{(2)}_i}{N},
\qquad 40\le N\le75,
\label{eq:fits}
\end{equation}
with the coefficients given in Table~\ref{tab:resid}. These are interpolation coefficients for the stated range and differ from the asymptotic ones of Eq.~\eqref{eq:asymptotic}. The difference is largest in the $b_G$ channel, where the ansatz in Eq.~\eqref{eq:fits} has no $\ln^{2}N$ term and the $n^{(2)}_i/N$ term amounts to $75\%$ of the other two at $N_*=55$. In the $\gamma_M$ channel, that term is $6\%$. At the pivot scale, the linear-response coefficients read
\begin{equation}
\begin{aligned}
\frac{\Delta n_s}{\theta_i}\bigg|_{N_*=55}&=\left(-5.15,\,-12.1,\,+25.8,\,-15.1\right)\times10^{-4},\\
\frac{\Delta r/r}{\theta_i}\bigg|_{N_*=55}&=\left(+0.041,\,+0.163,\,-0.181,\,+0.109\right),
\end{aligned}
\label{eq:pivotcoeffs}
\end{equation}
in the order of Eq.~\eqref{eq:residual}. We notice that the relative response of the tensor amplitude is about one hundred times larger than the shift of $n_s$, which is still far below observational reach for the matched parameters. More importantly, $\hat B_\Lambda$ is the only channel in which a positive parameter shifts $n_s$ toward the ACT value. According to Eq.~\eqref{eq:BGmass}, $\hat B_\Lambda$ is positive when the zeros of the asymptotic polynomials are real. In the direction $\gamma_M=b_G$ selected by Eq.~\eqref{eq:Kfixed}, the shift is $\Delta n_s=-1.73\times10^{-3}\,b_G$. The rescaled responses against $\ln N$ are displayed in Fig.~\ref{fig:resid}.

\begin{table}[t]
\centering
\begin{tabular}{lccccccl}
\toprule
$\theta_i$ & $n^{(0)}_i$ & $n^{(1)}_i$ & $n^{(2)}_i$ &
$r^{(0)}_i$ & $r^{(1)}_i$ & $r^{(2)}_i$ & class\\
\midrule
$\gamma_M$ & $+1.609$ & $-0.7675$ & $-5.03$ &
 $-0.923$ & $+0.7819$ & $+2.07$ & R, constant $\Rc$\\
$b_G$ & $+16.82$ & $-4.721$ & $-86.6$ &
 $-18.43$ & $+6.447$ & $+84.9$ & R $+$ S (exponent shift)\\
$\hat B_\Lambda$ & $-2.016$ & $+2.423$ & $+5.78$ &
 $-0.268$ & $-2.430$ & $+2.41$ & R, with $\ord(u)$ variation\\
$\hat\rho_\Lambda$ & $+1.871$ & $-1.582$ & $-5.97$ &
 $-0.379$ & $+1.592$ & $+0.74$ & R, with $\ord(u)$ variation\\
\bottomrule
\end{tabular}
\caption{Coefficients of the interpolating fits \eqref{eq:fits} over $40\le N\le75$. They are least-squares fits with relative weights to the linear response \eqref{eq:response} on the grid $N=40,45,\dots,75$. The maximal relative residuals, defined as $\max_N|{\rm fit}/{\rm response}-1|$, are $2.5\times10^{-5}$, $7.2\times10^{-4}$, $6\times10^{-6}$ and $3\times10^{-6}$ for $\nu_i$ and $5\times10^{-6}$, $2.6\times10^{-4}$, $1.5\times10^{-5}$ and $1.0\times10^{-5}$ for $\varrho_i$, in the order of the rows. The coefficients are strongly correlated, and their values depend on the grid at the level of the last digit quoted. The last column gives the class of Sec.~\ref{sec:iff}.}
\label{tab:resid}
\end{table}

It is worth noting that removability protects the shape of the potential but does not constrain $N_*$, which is determined by the post-inflationary history. For the minimal gravitational decay channel \cite{Vilenkin1985,MijicMorrisSuen1986}, one has
\begin{equation}
\Gamma_{\rm d}=\frac{N_s\Ms^{3}}{192\pi\mpl^{2}}\bigg|_{N_s=1}=7.3\ {\rm GeV},
\qquad
\Trh=\left(\frac{90}{\pi^{2}g_*}\right)^{1/4}\sqrt{\Gamma_{\rm d}\mpl}
=2.3\times10^{9}\ {\rm GeV},
\label{eq:Gamma}
\end{equation}
where $\Gamma_{\rm d}$ is the decay rate of the scalaron, $N_s$ is the number of scalar species in the final state, $\Trh$ is the reheating temperature, and $g_*=106.75$ is the number of relativistic degrees of freedom. Here the multiplicity $N_s$ cancels from $\delta N_*$, which depends only on the logarithmic derivative of $\Gamma_{\rm d}$. For matter-dominated oscillations, fixed $g_*$ and fixed reference $\mpl$, the matching of the pivot scale \cite{LiddleLeach2003} contains
\begin{equation}
N_*={\rm const}+\frac12\ln V_*-\frac13\ln\rho_{\rm end}+\frac1{12}\ln\rho_{\rm rh} ,
\label{eq:Nstar}
\end{equation}
where $V_*$ is the potential at the horizon exit of the pivot scale, and $\rho_{\rm end}$ and $\rho_{\rm rh}$ are the energy densities at the end of inflation and at reheating, respectively. All three terms respond to the running. The last one varies through the decay rate. Combining $\rho_{\rm rh}\propto\Trh^{4}$, $\Trh\propto\sqrt{\Gamma_{\rm d}\mpl}$, and $\Gamma_{\rm d}\propto\Ms^{3}/\mpl^{2}$, we obtain
\begin{equation}
\delta N_*^{(\Gamma_{\rm d}})=\left(\frac{\gamma_M}{4}-\frac{b_G}{12}\right)\ell_{\rm rh}
=0.176\,\gamma_M-0.059\,b_G,
\label{eq:reh}
\end{equation}
where $\ell_{\rm rh}=\ln(\Ms/H_*)=\ln[2/(1-u_0)]=0.706$. This logarithmic interval is narrow because the couplings entering $\Gamma_{\rm d}$ renormalize at the mass of the decaying particle. The other two terms of Eq.~\eqref{eq:Nstar} are evaluated at first order with the scalar amplitude held fixed. Then one has $\delta\ln V_*=\delta\ln\eps_*$ with $\eps_*\equiv\eps(N_*)$, the energy density at the end is $\rho_{\rm end}=\tfrac32V_{\rm end}$ with $V_{\rm end}/V_*$ taken from the numerical solution, and the reference value of $\Ms^{2}$ entering $\Gamma_{\rm d}$ is the one fixed by the amplitude. The results are listed in Table~\ref{tab:reh}, with $\Delta n_s^{(N_*)}=(\dd n_s/\dd N)\,\delta N_*$ and $\dd n_s/\dd N=6.21\times10^{-4}$ at $N_*=55$ from the tree-level numerical solution. The leading-order value $2/N_*^{2}$ is $6\%$ larger.

We find that the $e$-fold contribution opposes the shape shift in all four channels. It amounts to $29\%$ of the shape shift in the $\gamma_M$ channel, where the decay rate dominates, and to $14\%$ in the $\hat B_\Lambda$ channel, where the end of inflation dominates. In the direction $\gamma_M=b_G$, it is $\delta N_*=0.31\,b_G$ and $\Delta n_s^{(N_*)}=+1.9\times10^{-4}\,b_G$, which is $11\%$ of the shape shift $-1.73\times10^{-3}\,b_G$. However, three caveats apply. Firstly, the term $\rho_{\rm end}$ uses the improved potential at the end of inflation, where $\ell=-0.61$ and the first-order improvement is at the limit of its validity. Secondly, the decay rate is the tree-level width evaluated with running couplings, which is not a matched one-loop result. Thirdly, with the frame normalization of Sec.~\ref{sec:framenorm}, where $\rho_{\rm rh}/\mpl^{4}\propto(\Ms/\mpl)^{6}$, the decay-rate term becomes $\tfrac14(\gamma_M-b_G)\ell_{\rm rh}$ and vanishes in the direction $\gamma_M=b_G$ up to $\ord(10^{-11})$.

Reaching the ACT preferred value through this channel would require $\delta N_*\simeq15$. Even the extended choice $\ell_{\rm rh}=\ln(\Ms/\Trh)\simeq9.5$ leads to $\delta N_*^{(\Gamma_{\rm d})}\simeq2.4\gamma_M$, which is short by a factor of six even for $\gamma_M=\ord(1)$. Therefore, within this minimal decay model, the running cannot supply the required displacement. Note that this statement does not extend to other post-inflationary histories, which can change $N_*$ by several $e$-folds independently of the running \cite{StaroACTrefined}, e.g., through a stiff reheating phase \cite{Iacconi:2025odq}.

\begin{table}[t]
\centering
\begin{tabular}{lcccccc}
\toprule
$\theta_i$ & $\tfrac12\delta\ln V_*$ & $-\tfrac13\delta\ln\rho_{\rm end}$ & $\tfrac1{12}\delta\ln\rho_{\rm rh}$ & $\delta N_*$ & $\Delta n_s^{(N_*)}$ & fraction of shape\\
\midrule
$\gamma_M$ & $+0.020$ & $+0.039$ & $+0.184$ & $+0.243$ & $+1.51\times10^{-4}$ & $-29\%$\\
$b_G$ & $+0.081$ & $+0.006$ & $-0.021$ & $+0.067$ & $+4.2\times10^{-5}$ & $-3.4\%$\\
$\hat B_\Lambda$ & $-0.091$ & $-0.422$ & $-0.049$ & $-0.562$ & $-3.49\times10^{-4}$ & $-14\%$\\
$\hat\rho_\Lambda$ & $+0.055$ & $+0.213$ & $-0.222$ & $+0.045$ & $+2.8\times10^{-5}$ & $-1.9\%$\\
\bottomrule
\end{tabular}
\caption{First-order variation of the matching of the pivot scale in Eq.~\eqref{eq:Nstar} per unit running parameter at $N_*=55$, for the minimal gravitational decay channel with fixed $g_*$ and fixed scalar amplitude. The last column is $\Delta n_s^{(N_*)}$ divided by the shape coefficient of Eq.~\eqref{eq:pivotcoeffs}.}
\label{tab:reh}
\end{table}

\subsection{Ultraviolet--infrared matching}
\label{sec:matching}

The previous results hold for any completion whose local flow has the form of Eq.~\eqref{eq:linflow}, improved at the Einstein-frame Hubble rate. In order to estimate the size of the effect, we now match the coefficients of the asymptotic form factor to the nonlocality scale and propagate that estimate to $n_s$.

The framework connecting the form factor to the cosmological observables consists of the components summarized in Table~\ref{tab:dictionary}. The first one is the structural result of Sec.~\ref{sec:pfamily}. The second and third ones rest on the dimensional structure of the beta functions \cite{ModestoRachwalShapiro2018} combined with the lower bound $\Lstar\gtrsim1.2\times10^{14}$~GeV, which is of the order found in Ref.~\cite{KoshelevKumarMazumdar2016}. Note that this bound was obtained from the tensor spectrum of a class-J model and does not carry over to class~E, where the tensor spectrum is not dressed. We therefore use it as a nominal lower edge. The condition $H_*<\Lstar$ alone gives $\Lstar>1.5\times10^{13}$~GeV. The conclusions of Secs.~\ref{sec:iff}--\ref{sec:analytic} are independent of these assumptions. With no hierarchies among the polynomial coefficients, the single scale in the form factor dictates the ratios entering Eq.~\eqref{eq:BGgen}, and we obtain
\begin{equation}
b_G\simeq0.013\,\kappa_G\left(\frac{\Lstar}{\mpl}\right)^{2},
\qquad
\hat B_\Lambda\simeq5.7\times10^{7}\kappa_\Lambda
\left(\frac{\Lstar}{\mpl}\right)^{4},
\qquad
|\kappa_G|,|\kappa_\Lambda|=\ord(1).
\label{eq:bGmatch}
\end{equation}
According to Eq.~\eqref{eq:BGmass}, the dimensionless factors are given by
\begin{equation}
\kappa_G=\frac{1}{6\Lstar^{2}}\bigg[5\sum_i m_{i,2}^{2}+\sum_i m_{i,0}^{2}\bigg],
\qquad
\kappa_\Lambda=\frac{1}{2\Lstar^{4}}\bigg[5\sum_i m_{i,2}^{4}+\sum_i m_{i,0}^{4}\bigg],
\label{eq:kappas}
\end{equation}
and the prefactors of Eq.~\eqref{eq:bGmatch} are $2/(4\pi)^{2}=0.0127$ and $1/(4\pi)^{2}\hat V_0$ with $\hat V_0=V_0/\mpl^{4}=1.12\times10^{-10}$. Here the signs of $\kappa_G$ and $\kappa_\Lambda$ are those of the sums over zeros. For $\Npol$ real zeros at $m_i^{2}=\Lstar^{2}$ in each sector, one has $\kappa_G=\Npol$ and $\kappa_\Lambda=3\Npol$. For the equal ratios of Sec.~\ref{sec:rg}, one has $\kappa_G=1$ and $\kappa_\Lambda=-3$. Moreover, for a monomial $P(z)$, as used in Sec.~\ref{sec:poles} to evaluate $c_*$, one finds $\kappa_G=\kappa_\Lambda=0$ and the minimal-subtraction running of $\mpl^{2}$ and $\rho_\Lambda$ vanishes. A nonzero effect therefore requires subleading terms of the asymptotic polynomial with coefficients of order one in units of $\Lstar$, which we assume in what follows. Requiring the running vacuum energy not to exceed the plateau energy, i.e., $|\hat B_\Lambda|\ln(\Lstar/H_*)<1$, is a naturalness condition and not a derived bound. For $\kappa_\Lambda=1$, it restricts the window to
\begin{equation}
1.2\times10^{14}\ {\rm GeV}\lesssim\Lstar\lesssim1.7\times10^{16}\ {\rm GeV}.
\label{eq:window_L}
\end{equation}
At the upper boundary, $\hat B_\Lambda=1/\ln(\Lstar/H_*)\simeq0.141$ holds by construction and depends on $\kappa_\Lambda$ only through the logarithm. For $|\kappa_\Lambda|=3$, the boundary moves to $1.3\times10^{16}$~GeV and $|\hat B_\Lambda|=0.147$, and for $\kappa_\Lambda=9$ to $1.0\times10^{16}$~GeV and $\hat B_\Lambda=0.153$. By using the linear response in Eq.~\eqref{eq:pivotcoeffs}, we then obtain
\begin{equation}
|\Delta n_s|^{\rm lin}_{\rm max}
=\frac{2.58\times10^{-3}}{\ln(\Lstar/H_*)}=3.6\times10^{-4},
\qquad
\Delta n_s^{\rm num}=+2.6\times10^{-4}\quad(\hat B_\Lambda=0.141) ,
\label{eq:maxshift}
\end{equation}
where the second value is the numerical solution for the full deformation at that amplitude. The two values differ by $40\%$ because $\hat B_\Lambda=0.14$ is outside the linear regime, and the linear value is an upper estimate. In addition, the exact background evolution with second-order spectral formulae (see Sec.~\ref{sec:limits}) gives $+2.9\times10^{-4}$, and the $e$-fold channel of Table~\ref{tab:reh} lowers the estimate by $14\%$. The $\mpl$ and $\Ms$ channels contribute $-1.1\times10^{-9}$ at that boundary, as detailed in Table~\ref{tab:scan}. The central value favored by ACT requires $\Delta n_s=+9.3\times10^{-3}$, which is larger by a factor between $25$ and $35$ depending on which of these evaluations is used.

However, five qualifications apply. Firstly, the estimate is conditional on the naturalness condition above, which is neither a small-expansion bound at the matching scale nor a statement valid for arbitrary boundary conditions and form factors. Secondly, it assumes the renormalization condition $\hat\rho_\Lambda(\mu_0)=0$ at the pivot scale, and this is a choice. A rescaling $\mu=cH$ of the improvement scale at fixed reference couplings shifts $\ell$ by $\ln c$, which is equivalent to $\hat\rho_\Lambda\to\hat\rho_\Lambda+\hat B_\Lambda\ln c$. Therefore, only the combination $\Delta n_s=(25.8\,\hat B_\Lambda-15.1\,\hat\rho_\Lambda)\times10^{-4}$ with a stated condition on $\hat\rho_\Lambda$ is meaningful, and for $c=2$ at fixed $\hat\rho_\Lambda$ the response to $\hat B_\Lambda$ changes by $-41\%$. If $|\hat\rho_\Lambda(\mu_0)|$ is allowed the same range $0.14$ as $\hat B_\Lambda$, the linear estimate becomes $|\Delta n_s|\le5.8\times10^{-4}$, i.e., a factor of $16$ below the ACT displacement. Note that a vacuum energy at the pivot scale comparable to the plateau energy lies outside the regime in which the deformation is a small correction. Thirdly, the largest shift is reached only for $\hat B_\Lambda>0$, i.e., for $\kappa_\Lambda>0$ in Eq.~\eqref{eq:kappas}. For $\hat B_\Lambda<0$, the shift is negative, and for $\hat B_\Lambda\lesssim-0.07$ the improved potential never reaches $\eps_V=1$, because the vacuum-energy term dominates near the minimum, where $|\ell|$ is large and the improvement is not valid. Fourthly, across the whole window one has $\hat B_\Lambda/b_G=4.5\times10^{9}(\Lstar/\mpl)^{2}>10$. This means that the estimate is a statement about the running of the vacuum energy and not about the running of $\mpl$ or $\Ms$. Fifthly, the coefficients $b_G\propto\Lstar^{2}$ and $B_\Lambda\propto\Lstar^{4}$ are minimal-subtraction coefficients generated by modes of virtuality $\Lstar$, whereas $H_*<\Lstar$ throughout the window. In a mass-dependent scheme, the dependence of low-energy quantities on $H$ is expected to be reduced below such a threshold \cite{GorbarShapiro2003}. Using these coefficients at $\mu=H$ is an assumption, and Eq.~\eqref{eq:maxshift} should be read as an upper estimate within it.

The last component of Table~\ref{tab:dictionary} operates in the reverse direction. In class~J, the coefficient of $R^{2}$ is the lowest polynomial coefficient of the form factor. Since the completed theory inherits the classical background exactly, it is fixed by the scalar amplitude as
\begin{equation}
\omega_{0,R}=\alpha_R^{\rm tree}=\frac{\mpl^{2}}{12\Ms^{2}}
{}=\frac{(1-u_0)^{4}}{512\pi^{2}u_0^{2}A_s}=5.6\times10^{8}.
\label{eq:omega0Rvalue}
\end{equation}
The leading-order expression $N_*^{2}/288\pi^{2}A_s=5.1\times10^{8}$ is $10\%$ lower. The uncertainty budget is dominated by the reference scale and not by the scalar amplitude, which is known at the percent level. Since $\omega_{0,R}\propto N_*^{2}$ at leading order, a shift $\delta N_*=1$ alters the result by $2/N_*\simeq3.6$ percent. In addition, Eq.~\eqref{eq:omega0Rvalue} shows how $\omega_{0,R}$ compares with the value $\mpl^{2}/\Lstar^{2}$ expected for a form factor with a single scale, which ranges from $4\times10^{8}$ to $2\times10^{4}$ across the window in Eq.~\eqref{eq:window_L}. The two agree for $\Lstar\simeq\sqrt{12}\,\Ms=1.0\times10^{14}$~GeV, which is close to the lower edge. For larger $\Lstar$, the $R^{2}$ coefficient exceeds the single-scale expectation by $\Lstar^{2}/12\Ms^{2}$, up to $3\times10^{4}$. Whether such a hierarchy is natural within a concrete completion is beyond the scope of the present work.

\begin{table}[t]
\centering
\begin{tabular}{lll}
\toprule
ultraviolet datum & infrared consequence & status \\
\midrule
loop order of divergences ($\Npol\ge3$)
 & local beta functions one-loop exact
 & Sec.~\ref{sec:pfamily}\\
$\omega_{\Npol-1,i}/\omega_{\Npol,i}\sim\Lstar^{2}$
 & $b_G\simeq0.013\,\kappa_G(\Lstar/\mpl)^{2}$
 & dimensional\\
$\omega_{\Npol-2,i}/\omega_{\Npol,i}\sim\Lstar^{4}$
 & $\hat B_\Lambda\simeq5.7\times10^{7}\kappa_\Lambda(\Lstar/\mpl)^{4}$
 & dimensional\\
$b_G,\hat B_\Lambda$ with Eq.~\eqref{eq:pivotcoeffs}
 & $|\Delta n_s|{}\simeq3\times10^{-4}\ \text{at most}$
 & estimate\\
$\omega_{0,R}$, lowest coefficient
 & $\Ms^{2}=\mpl^{2}/12\omega_{0,R}$, fixed by $A_s$
 & measurement\\
\bottomrule
\end{tabular}
\caption{The ultraviolet--infrared dictionary. $\Npol$ denotes the degree of the
asymptotic polynomial. The second, third and fifth rows use ingredients of class~J.}
\label{tab:dictionary}
\end{table}

\begin{table}[t]
\centering
\begin{tabular}{ccccc}
\toprule
$\Lstar/\mpl$ & $\Lstar$ [GeV] & $b_G$ & $\hat B_\Lambda$ & $\Delta n_s$\\
\midrule
$10^{-5}$ & $2.4\times10^{13}$ & $1.3\times10^{-12}$ & $5.7\times10^{-13}$ & $-7\times10^{-16}$\\
$10^{-4}$ & $2.4\times10^{14}$ & $1.3\times10^{-10}$ & $5.7\times10^{-9}$ & $+1.4\times10^{-11}$\\
$10^{-3}$ & $2.4\times10^{15}$ & $1.3\times10^{-8}$ & $5.7\times10^{-5}$ & $+1.5\times10^{-7}$\\
$7.07\times10^{-3}$ & $1.7\times10^{16}$ & $6.3\times10^{-7}$ & $0.141$ & $+3.6\times10^{-4}$ ($+2.6\times10^{-4}$ numerical)\\
\bottomrule
\end{tabular}
\caption{Matched running parameters and induced shifts for
$\kappa_G=\kappa_\Lambda=1$, $\gamma_M=b_G$, $\hat\rho_\Lambda=0$ and $N_*=55$, from the linear response \eqref{eq:pivotcoeffs}. The first row lies below the
lower bound of Ref.~\cite{KoshelevKumarMazumdar2016} and shows the scaling only. There the two contributions are comparable and the sign is that of the $b_G$ term.
At the last row $\Delta n_s$ is independent of $\kappa_\Lambda$.}
\label{tab:scan}
\end{table}

\begin{figure}[t]
\centering
\includegraphics[width=0.95\textwidth]{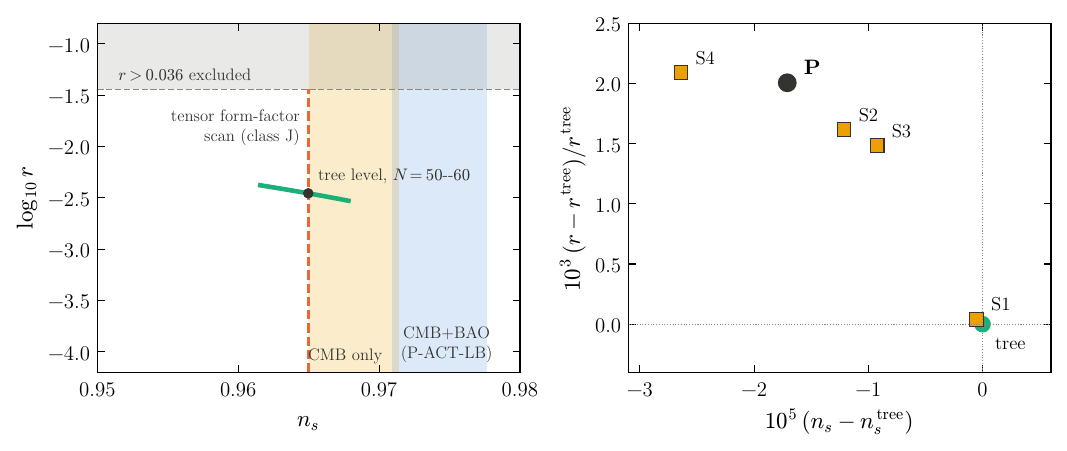}
\caption{Left: the $(n_s,r)$ plane with the tree-level trajectory for $N\in[50,60]$ and the range of $r$ accessible at fixed $n_s$ through the tree-level tensor form factor in class~J. The vertical bands are the $68\%$ intervals $n_s=0.9743\pm0.0034$ of Ref.~\cite{ACTDR6}, for ACT and Planck with CMB lensing and DESI DR1 BAO, and $n_s=0.9682\pm0.0032$ of Ref.~\cite{Balkenhol2025}, for CMB data alone. The horizontal line is the $95\%$ upper limit $r<0.036$ of Ref.~\cite{BICEPKeck2021}. These are one-dimensional marginal constraints obtained with different data sets and assumptions. They are drawn for scale and do not represent a joint likelihood in the $(n_s,r)$ plane. Right: the renormalization-group
displacement at $N_*=55$ for illustrative amplitudes. The squares switch on $\gamma_M=10^{-3}$ (S1), $b_G=10^{-2}$ (S2), $(\gamma_M,b_G,\hat B_\Lambda)=(2,1,0.5)\times10^{-2}$ (S3) and $\gamma_M=5.5\times10^{-2}$ (S4). They are decompositions and not scenarios, since
Eq.~\eqref{eq:Kfixed} locks $\gamma_M$ to $b_G$. The black point P is the
direction $\gamma_M=b_G=10^{-2}$, giving
$\Delta n_s=-1.71\times10^{-5}$. The matched values of $\gamma_M$ and $b_G$ are at least four orders of magnitude smaller than those shown, whereas $\hat B_\Lambda$ may reach $0.14$ (Table~\ref{tab:scan}).}
\label{fig:nsr}
\end{figure}

\section{Propagator poles and the tensor sector: perturbative estimates}
\label{sec:poles}

The scalar potential analysis does not by itself control the pole structure or the tensor amplitude. These quantities probe the dressed inverse propagator and are checked separately in this section. It is worth noting that the estimates given here are conditional. They use the flat-space propagator on the real momentum axis and, for the tensor sector, class-J ingredients whose relation to the action used for the running has not been derived.

An improved propagator can develop poles absent from the tree-level theory, and the perturbative treatment remains valid only where such poles do not appear. It was argued in Ref.~\cite{ModestoOrlando2026} that loop-induced extra poles arise only when the perturbative corrections become comparable to the classical terms. Here we quantify that observation. By using the normalizations of Eq.~\eqref{eq:spin2norm} and $\delta\alpha_R=\alpha_R-\alpha_R^{\rm tree}$, which prevents double counting of the scalaron pole, the spin-2 inverse propagator takes the form
\begin{equation}
\Pi_2^{-1}=\frac{\mpl^{2}}{2}k^{2}e^{\Hf_2}
\left[1-\frac{4k^{2}}{\mpl^{2}}e^{-\Hf_2}
\left(\alpha_C+B_C\ln\frac{|k^{2}|}{\mu^{2}}\right)+\cdots\right],
\label{eq:Pi2}
\end{equation}
and an analogous expression holds in the spin-0 sector with $4\to12$ and an additional factor $(1+k^{2}/\Ms^{2})^{-1}$, where $\Hf_j$ with $j=0,2$ denotes the form-factor exponent in the spin-$j$ sector. At momenta relevant to inflation, both bracketed terms deviate from unity by $\ord(10^{-10})$ for couplings of order one, and hence no additional pole appears in that regime. An extra zero requires $e^{\Hf_{j}(y)}/y=(4\Lstar^{2}/\mpl^{2})(\alpha_C+\cdots)$ with $y=|k^{2}|/\Lstar^{2}$. The left side has a strict lower bound, and its minimum, which is attained at $y=y_*$,
\begin{equation}
c_*\equiv\min_{y>0}\frac{e^{\Hf_{j}(y)}}{y}
\label{eq:cstar}
\end{equation}
determines the safe operational window as
\begin{equation}
\left|\alpha_C+B_C\ln\frac{\Lstar^{2}}{\mu^{2}}\right|
\lesssim c_*\frac{\mpl^{2}}{4\Lstar^{2}},
\qquad
\left|\delta\alpha_R+B_R\ln\frac{\Lstar^{2}}{\mu^{2}}\right|
\lesssim c_*\frac{\mpl^{2}}{12\Lstar^{2}} .
\label{eq:window}
\end{equation}
The logarithm in Eq.~\eqref{eq:Pi2} depends on the momentum, i.e., $\ln(|k^{2}|/\mu^{2})=\ln y+\ln(\Lstar^{2}/\mu^{2})$, and Eq.~\eqref{eq:window} evaluates it at $y=1$. A sufficient condition for the absence of a zero on the real axis is given by
\begin{equation}
\sup_{y>0}\;\frac{4\Lstar^{2}}{\mpl^{2}}\;y\,e^{-\Hf_2(y)}\left|\alpha_C+B_C\ln\frac{y\Lstar^{2}}{\mu^{2}}\right|<1 ,
\label{eq:supcond}
\end{equation}
which is implied by
\begin{equation}
\left|\alpha_C+B_C\ln\frac{\Lstar^{2}}{\mu^{2}}\right|+c_*s_*\left|B_C\right|<c_*\frac{\mpl^{2}}{4\Lstar^{2}},
\qquad
s_*\equiv\sup_{y>0}\;y\,e^{-\Hf_2(y)}\left|\ln y\right| .
\label{eq:supcond2}
\end{equation}
For the form factors below, we find $s_*=0.36$ to $0.37$. Therefore, the momentum dependence of the logarithm changes the right-hand side of Eq.~\eqref{eq:window} by $c_*s_*|B_C|\simeq0.04$ for the value of $B_C$ in Eq.~\eqref{eq:dnTalphaC}, which is negligible against values above $10^{3}$. The spin-0 condition has the same form away from the scalaron pole, whose displacement is given in Eq.~\eqref{eq:massshift}. These are statements about the real axis. A minimum on the real positive axis does not exclude complex zeros, and $\ln|k^{2}|$ does not define the analytic continuation, branch cut or causal prescription that a statement about complex poles would require. We refer the readers to Ref.~\cite{Shapiro2015} for a discussion of loop-induced complex poles in nonlocal gravity.

\begin{figure}[t]
\centering
\includegraphics[width=0.95\textwidth]{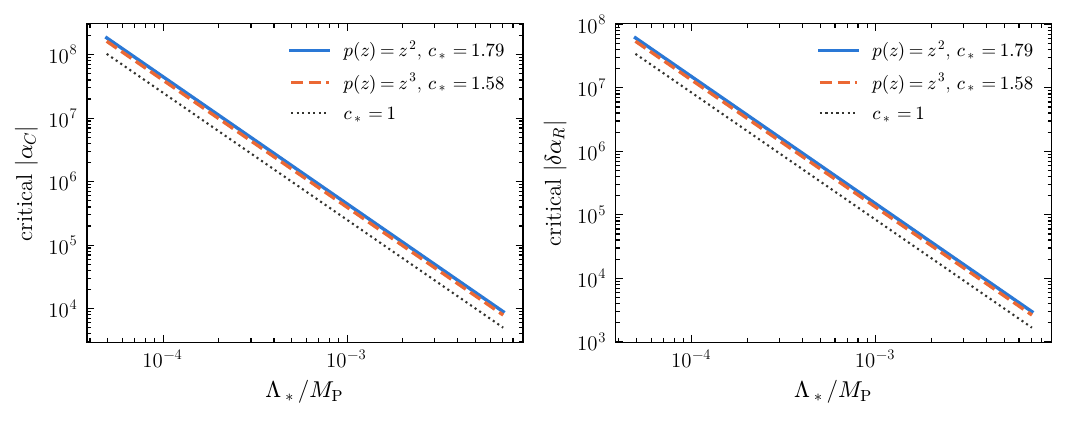}
\caption{Critical couplings for an extra spin-2 (left) and spin-0 (right) zero
of the improved inverse propagator across the window \eqref{eq:window_L}. Solid
and dashed curves use $c_*=1.79$ and $c_*=1.58$, obtained from Eq.~\eqref{eq:formfactor} with $\alpha_{\Hf}=2$, and the dotted line is the naive
$c_*=1$. Any perturbative coupling lies far below all three.}
\label{fig:window}
\end{figure}

We evaluate $c_*$ for form factors of the type introduced in Ref.~\cite{Tomboulis1997}, in the one-parameter form used in Ref.~\cite{ModestoOrlando2026}. It reads
\begin{equation}
\Hf(z)=\frac{\alpha_{\Hf}}{2}\left[\gamma_E+\Gamma\!\left(0,P(z)^{2}\right)+\ln P(z)^{2}\right],
\qquad
e^{\Hf(z)}\xrightarrow[z\to\infty]{}e^{\alpha_{\Hf}}\gamma_E/2\,|P(z)|^{\alpha_{\Hf}},
\label{eq:formfactor}
\end{equation}
where $\Gamma(0,x)$ is the incomplete gamma function, $\alpha_{\Hf}$ is a positive parameter, and $P$ is a real polynomial with $P(0)=0$. For $\alpha_{\Hf}=2$, the choices $P(z)=z^{2}$ and $P(z)=z^{3}$ give $c_*=1.79$ at $y_*=0.73$ and $c_*=1.58$ at $y_*=0.75$, respectively. For $\alpha_{\Hf}=1$ and $P(z)=z^{4}$, one finds $c_*=1.34$ at $y_*=0.86$. These cases have $e^{\Hf}\sim z^{4}$ or $z^{6}$ and satisfy the condition $\gamma_{\rm uv}>2$ of Sec.~\ref{sec:pfamily}. On the other hand, the choice $\alpha_{\Hf}=2$, $P(z)=z$ gives $c_*=2.17$ at $y_*=0.83$ but has $e^{\Hf}\sim z^{2}$ and does not satisfy it. In the normalization of Ref.~\cite{ModestoRachwalShapiro2018}, which is Eq.~\eqref{eq:formfactor} with $\alpha_{\Hf}=1$, the choice $P(z)=z^{3}$ gives $c_*=1.40$ at $y_*=0.86$, while $P(z)=z$ has no finite minimum, the infimum $e^{\gamma_E/2}=1.33$ being approached as $y\to\infty$. This means that the value of $c_*$ depends on the normalization and on the UV power of the form factor. The safe window is therefore broader than the naive estimate $c_*=1$ by a factor between $1.3$ and $1.8$. Since $\mpl^{2}/\Lstar^{2}$ is bounded below by $2.0\times10^{4}$ throughout the region in Eq.~\eqref{eq:window_L}, the critical couplings exceed $10^{3}$, as illustrated in Fig.~\ref{fig:window}. This is consistent with the statement of Ref.~\cite{ModestoOrlando2026}. These bounds are perturbative conditions and not a nonperturbative proof of unitarity. In addition, the scalaron pole itself experiences a finite displacement,
\begin{equation}
\delta\Ms^{2}[k=-\frac{12\Ms^{4}}{\mpl^{2}}e^{-\Hf_0(\Ms^{2}/\Lstar^{2})}
\left[\delta\alpha_R+B_R\ln\frac{\Ms^{2}}{\mu^{2}}\right],
\qquad
\left|\frac{\delta\Ms^{2}}{\Ms^{2}}\right|=\ord(10^{-9}).
\label{eq:massshift}
\end{equation}

Note that Eqs.~\eqref{eq:cstar}--\eqref{eq:window} assume a flat spacetime background, whereas the physical background is quasi-de Sitter. There are two separate issues. The first one is kinematical. In the transverse-traceless sector on a maximally symmetric background, the appropriate massless operator is $\Boxop-2H^{2}$ and not $\Boxop$, because $R_{\mu\alpha\nu\beta}h^{\alpha\beta}=-H^{2}h_{\mu\nu}$ for traceless $h_{\mu\nu}$. If the spin-2 kernel depends on this operator alone, its argument is shifted to
\begin{equation}
\tilde y=\frac{-k^{2}+2H^{2}}{\Lstar^{2}} ,
\label{eq:ydS}
\end{equation}
and the shift $2H_*^{2}/\Lstar^{2}$ is $3.0\times10^{-2}$ at the lower boundary of Eq.~\eqref{eq:window_L} and $1.5\times10^{-6}$ at the upper boundary, which is small compared with $y_*\simeq0.8$. However, this observation is not a de Sitter verification of the estimate. We have not carried out the operator and mode analysis on the inflationary background, and a one-dimensional flat-space minimization with a shifted argument cannot replace it. The second issue is dynamical and lies outside the present truncation. It was shown in Ref.~\cite{Tokareva2024} that a no-ghost condition imposed in flat space does not carry over to de Sitter space, where the graviton two-point function of an infinite-derivative theory can have an infinite tower of complex-conjugate poles, while the ghost-free condition was derived directly from a de Sitter background expansion in Ref.~\cite{FengYang2026}. These states are intrinsic to the tree-level theory on the inflationary background. They are not generated by the running, and their classical stability corresponds to the condition stated in Sec.~\ref{sec:limits} \cite{Addazi2024}. We therefore read Eqs.~\eqref{eq:cstar}--\eqref{eq:window} as a condition on \emph{loop-induced additional} poles on top of the intrinsic tree-level spectrum, and not as a statement about de Sitter unitarity.

We now turn to the tensor sector. Removability concerns the scalar potential and gives no protection to the tensor sector. Here the situation differs between the two classes of Sec.~\ref{sec:classes}. In class~E, the tree-level tensor spectrum is that of the local model. In class~J, the Weyl form factor dresses the tensor spectrum already at tree level and introduces a free function \cite{Koshelev2016,KoshelevKumarStarobinsky2023}. We write the class-J result with a function $\omega_T$, and class~E corresponds to $\omega_T\equiv0$, viz.,
\begin{equation}
r_{\rm RG}=16\eps_V^{\rm RG}\,e^{2\omega_T(\bar R/6\Mnl^{2})}
\left[1+\delta_T^{\log}\right],
\qquad
\left|\delta_T^{\log}\right|
=\frac{4H^{2}}{\mpl^{2}}e^{-\Hf_2}
\left|\alpha_C+B_C\ln\frac{H^{2}}{\mu^{2}}\right|=\ord(10^{-10}),
\label{eq:rRG}
\end{equation}
where $\bar R=12H^{2}$ is the background curvature and $\Mnl$ is the scale of the Weyl form factor, which need not coincide with $\Lstar$. As in the tree-level analysis, the freedom in $\omega_T$ governs the amplitude of $r$ \cite{KoshelevKumarStarobinsky2023}. In Fig.~\ref{fig:nsr}, we place the three effects on a single set of axes, namely the tree-level trajectory, the scan in $\omega_T$ that varies $r$ by orders of magnitude at fixed $n_s$, and the RG displacement, which is invisible without magnification.

There are two loop-level mechanisms acting on the tensor observables. The first one is the running of $\alpha_C$ within the dressed inverse propagator. The improvement at $\mu=H(\phi)$ promotes $\alpha_C$ to a field-dependent quantity through $\ell$, and the bracketed term in Eq.~\eqref{eq:rRG} acquires the field dependence as
\begin{equation}
\delta_T^{\log}(\phi)=-\frac{4H^{2}(\phi)}{\mpl^{2}}\,e^{-\Hf_2}
\left[\alpha_C^{(0)}+B_C\,\ell(\phi)\right].
\label{eq:dTfield}
\end{equation}
By using $\dd\ell/\dd\ln k=-\eps$, where $k$ here denotes the comoving wavenumber, the induced contribution to the tensor tilt $n_T$ is given by
\begin{equation}
\Delta n_T^{(\alpha_C)}
=\frac{4\eps H^{2}}{\mpl^{2}}e^{-\Hf_2}B_C
=3.2\times10^{-14}\,B_C,
\qquad
 B_C=\frac{2\chi/9+397/40}{(4\pi)^{2}}\bigg|_{\chi=1}=6.4\times10^{-2},
\label{eq:dnTalphaC}
\end{equation}
evaluated at $N_*=55$, where $\eps=2.19\times10^{-4}$ and $H_*^{2}/\mpl^{2}=3.6\times10^{-11}$. Here the value of $B_C$ is that of the six-derivative theory with $\chi\equiv\omega_{1,C}/\omega_{1,R}$ \cite{RachwalModestoPinzulShapiro2021}, which we use at $\chi=1$ as an order-of-magnitude proxy, since the corresponding coefficient for $\Npol\ge3$ is not needed at higher accuracy. The ratio $\chi$ and the applicability of this coefficient to the completion are not derived, and the result should be read in the parametric form $3.2\times10^{-14}B_C$, which is $2.0\times10^{-15}$ for the value above. Against the tree-level expectation $n_T=-2\eps=-4.4\times10^{-4}$, this is a relative correction of $5\times10^{-12}$.

The second mechanism exists in class~J only and is the deformation of the form-factor argument. The dressing factor in Eq.~\eqref{eq:rRG} is evaluated at $x_T=\bar R/6\Mnl^{2}$, and the scale $\Mnl$ does not run for the same reason that $\Lstar$ remains constant. The tensor spectrum is then proportional to $H^{2}e^{2\omega_T(x_T)}$ with $x_T\propto H^{2}$, and the tree-level tilt reads
\begin{equation}
n_T=-2\eps\left[1+2x_T\,\omega_T'(x_T)\right] ,
\label{eq:nTtree}
\end{equation}
which differs from $-2\eps$ when $x_T\omega_T'=\ord(1)$, where $\omega_T'\equiv\dd\omega_T/\dd x_T$. Under the deformation, we vary the whole spectrum at fixed $N$ and fixed scalar amplitude. If the scalar spectrum has the standard form, one has $H_*^{2}\propto\eps_*$ and therefore $\delta\ln x_T=\delta\ln\eps$, which gives
\begin{equation}
\begin{aligned}
\frac{\Delta r}{r}&=\left[1+2x_T\,\omega_T'\right]\delta\ln\eps,\\
\frac{\Delta n_T}{n_T}&=\left[1+\frac{2x_T\left(\omega_T'+x_T\,\omega_T''\right)}{1+2x_T\,\omega_T'}\right]\delta\ln\eps ,
\end{aligned}
\label{eq:dnTff}
\end{equation}
with $\delta\ln\eps$ given by Eq.~\eqref{eq:response}. This means that the amplitude of $r$ is not invariant at the pivot scale. Only the explicit dependence of the dressing on $\ell$ vanishes there, while the slow-roll factor changes at fixed $N$, as shown by the values of $\Delta r/r$ in Eq.~\eqref{eq:pivotcoeffs}. Both relative changes are of order $\delta\ln\eps=\ord(K\ln N/N)$ and not $\ord(K)$. Here the fixed-field variation $\delta x_T/x_T=\Delta$ is compensated by the displacement of the $e$-fold map, exactly as for $n_s$. At the upper boundary of Eq.~\eqref{eq:window_L}, one has $\delta\ln\eps=-1.9\%$, and hence the changes are at the percent level times factors of order one that depend on $\omega_T$. This is a structural property and not an observational prediction, because $|n_T|$ is itself $\ord(10^{-4})$ and far below any projected sensitivity. The function $\omega_T$ dictates $r$ at tree level, and the running modifies $r$ and $n_T$ only through $\delta\ln\eps$. Therefore, the primordial gravitational-wave program of Ref.~\cite{CalcagniModesto2024} probes the tree-level structure of the completion and not its loop corrections.

\section{Discussion}
\label{sec:discussion}

\subsection{Field logarithms, potential logarithms, and the proposals in the literature}
\label{sec:classification}

The criterion evaluates a proposed correction through the classes of Sec.~\ref{sec:iff}, without a computation of the observables. For the deformations of primary interest, $\Rc$ grows with $N$. A field logarithm gives $\Rc=\ord(N/\ln N)$ and hence $\Delta n_s=\ord(1/\ln N)$ per unit amplitude, which is nearly independent of $N$. On the other hand, a polynomial in the field gives $\Rc=\ord(N\ln N)$ and a growing shift $\Delta n_s=\ord(\ln N)$. Therefore, non-removable deformations violate the attractor behavior more strongly than a $1/N$ counting argument would suggest.

According to Eq.~\eqref{eq:ell}, setting the improvement scale to the Einstein-frame Hubble rate makes the RG time a logarithm of the potential. As a result, the first-order deformation is affine in $\ln\Vst$ and satisfies Eq.~\eqref{eq:Rtest}. Note that this holds for affine functions of $\ln\Vst$ only. In general, one has $\Rc[F(V)]=VF'(V)$, which is constant only for $F=c_1+c_2\ln V$, and the $\ell^{2}$ term of Eq.~\eqref{eq:product} is a function of $\Vst$ with nonconstant $\Rc$. On the other hand, matter loops generate logarithms of the field or of a field-dependent mass. A Coleman--Weinberg contribution from a field whose mass depends on the inflaton through a direct coupling introduces terms proportional to $m^{4}(\phi)\ln m^{2}(\phi)$. On a monotonic branch, any function of the field can be rewritten as a function of $\Vst$, and hence the distinction is not whether a correction depends on $\Vst$ alone. It lies in the functional dependence and in the scaling of the derivative. A logarithm of the field is not affine in $\ln\Vst$, and its $\Rc$ grows on the plateau. It is worth noting that whether a matter correction is removable is decided by its mass functions and couplings and not by its origin. The authors of Ref.~\cite{AttractorsRadACT} applied to $\alpha$-attractors a correction of the form $V\to V[1+\delta\ln(\phi/\mpl)]$ in the Einstein frame. This is the E-model case of that reference. Its corrections to $\xi$-attractors are defined in the Jordan frame and require the transformation to the Einstein frame and a canonical normalization before the test is applied. For this deformation on the Starobinsky plateau, we obtain
\begin{equation}
\Rc[\delta\ln(\phi/\mpl)]=\frac{\Vst}{\Vst'}\frac{\delta}{\phi}
=\delta\,\frac{1-u}{2au\phi}
\simeq\frac{\delta}{2u\ln(1/u)} ,
\label{eq:Rmatterlog}
\end{equation}
which diverges as $u\to0$. In Table~\ref{tab:test}, we apply the test to every deformation considered here and to those proposed in the recent literature.

It is worth noting that a field logarithm can also be induced gravitationally. The one-loop effective potential of scalar-tensor gravity around flat spacetime was derived in Ref.~\cite{Arbuzov:2020pgp}. For the nonminimal coupling $-\tfrac12\xi R\phi^{2}$, i.e., the last operator of Eq.~\eqref{eq:opbasis}, Eq.~(28) of that reference gives
\begin{equation}
V_{\rm AL}(\phi)=\frac{m_{\rm obs}^{2}}{3\kappa^{2}\xi^{2}}\ln\left(1+\frac32\,\xi^{2}\kappa^{2}\phi^{2}\right),
\label{eq:VAL}
\end{equation}
where $\kappa$ is the gravitational coupling of that reference and $m_{\rm obs}$ is the observed mass imposed by the renormalization prescription. If this contribution is added to the Starobinsky potential in the same canonical field, the deformation $\Delta=V_{\rm AL}/\Vst$ gives
\begin{equation}
\Rc[\Delta]=\frac{m_{\rm obs}^{2}}{2aV_0}\,\frac{\phi}{u(1-u)\left[1+\tfrac32\xi^{2}\kappa^{2}\phi^{2}\right]}-\Delta ,
\label{eq:RAL}
\end{equation}
which grows as $1/u$ on the plateau, and hence the deformation is not removable. For $\xi\kappa\phi\ll1$, Eq.~\eqref{eq:VAL} reduces to the mass term $\tfrac12m_{\rm obs}^{2}\phi^{2}$, which is of the polynomial form of Eq.~\eqref{eq:Rpoly} below with $d=2$. For $\xi\kappa\phi\gg1$, it becomes the field logarithm of Eq.~\eqref{eq:Rmatterlog} with $\delta=2m_{\rm obs}^{2}/3\kappa^{2}\xi^{2}V_0$. This example shows that a gravitationally induced contribution can also carry a non-removable field dependence, in line with the statement above that removability is decided by the functional form and not by the origin of a correction. However, three remarks are in order. Firstly, Eq.~\eqref{eq:VAL} assumes a renormalization prescription with a nonzero observed mass despite a vanishing tree-level mass. If a vanishing renormalized mass is imposed instead, this contribution vanishes, and the underlying scaleless integral is discarded in dimensional regularization. Secondly, the calculation is performed around flat spacetime. Thirdly, it has not been established that this contribution survives in the weakly nonlocal completions considered here with the improvement scale $\mu=H(\phi)$. We therefore do not include it in the deformation of Eq.~\eqref{eq:Delta}.

Two entries of the table calibrate the criterion. The first one is $\Delta=c\,u$. It passes the test, because the expansion $\ln\Vst=-2u+\ord(u^{2})$ exhibits it as a member of class~R. The second one is an exact constant shift of the inflaton, $\Vst(\phi+\phi_{\rm s})$. This is the $c_3$ direction of Eq.~\eqref{eq:invclass}, and it leaves both $n_s$ and $r$ invariant to machine precision for all tested amplitudes up to $\phi_{\rm s}=0.1\mpl$. Neither produces a strictly constant $\Rc$. The first one gives $-c(1-u)/2$ and the second one gives $-a\phi_{\rm s}/(1-u)$, both varying at $\ord(u)$. Between $N=40$ and $N=55$, each changes by $0.4$ percent. The vacuum-energy terms of the flow change by $0.9$ and $1.8$ percent over the same interval, whereas the four non-removable entries of the lower block change by $21$ to $45$ percent. We therefore apply the criterion uniformly at $\ord(u)$. At that precision, the test separates the two groups by more than a factor of ten.

Since the test involves a single differentiation, it applies to the deformations proposed since the ACT DR6 release without reproducing their analyses. For deformations specified in the Jordan frame, a compact transformation exists. Writing $f=f_0+\delta f$ with $f_0=R+R^{2}/6\Ms^{2}$ and varying $V=\tfrac12\mpl^{2}(Rf'-f)/f'^{2}$ at fixed $\phi$, and therefore at fixed $f'$, the two terms containing $\delta R$ cancel and one obtains
\begin{equation}
\Delta\equiv\frac{\delta V}{\Vst}
=-\frac{\delta f(R)}{Rf_0'-f_0}
=-\frac{6\Ms^{2}}{R^{2}}\,\delta f(R),
\qquad
R=3\Ms^{2}\,\frac{1-u}{u} ,
\label{eq:fRlemma}
\end{equation}
which converts any $f(R)$ proposal into an Einstein-frame deformation. For a generic monomial $\delta f=\lambda_n R^{n}/M^{2n-2}$, with a dimensionless coefficient $\lambda_n$ and a mass scale $M$, this produces $\Delta_{(n)}=-c_n[(1-u)/u]^{n-2}$ and leads to
\begin{equation}
\Rc\big[\Delta_{(n)}\big]
=-\frac{(n-2)\,c_n}{2}\frac{(1-u)^{n-2}}{u^{n-1}}
\simeq-\frac{(n-2)\,c_n}{2}\left(\frac{4N}{3}\right)^{n-1} ,
\label{eq:Rn}
\end{equation}
where $c_n=6\lambda_n\Ms^{2}(3\Ms^{2})^{n-2}/M^{2n-2}$. For $n=2$, the deformation is a constant, i.e., the amplitude direction $c_1$, because shifting the $R^{2}$ coefficient renormalizes $\Ms$. For $n=1$, it is $\Delta\propto u/(1-u)\propto\Vst'/\Vst$, i.e., the field-shift direction $c_3$, which corresponds to a rescaling of the Planck mass. However, every power $n\ge3$ is a non-removable deformation whose degree of violation grows with $n$. The companion formula for a deformation polynomial of degree $d$ in the canonical field reads
\begin{equation}
\Rc\big[c\,(a\phi)^{d}\big]
=c\,\frac{d(1-u)(a\phi)^{d-1}}{2u}
\simeq c\,\frac{2dN}{3}\left(\ln\frac{4N}{3}\right)^{d-1} ,
\label{eq:Rpoly}
\end{equation}
which reproduces the polynomial row of Table~\ref{tab:test} at $d=2$ and accommodates the marginal case at $d=1$. Applying Eq.~\eqref{eq:fRlemma} to a marginal deformation $R^{2}\to R^{2}(R/R_0)^{-2\alpha}$ with a reference curvature $R_0$, one finds $\Delta=2\alpha\ln(R/R_0)=2\alpha[a\phi+\ln(1-u)]+{\rm const}$ at first order in $\alpha$. Here the second component is removable, being equal to $\alpha\ln\Vst$ up to an additive constant, and gives $\Rc=\alpha$. The first component is the $d=1$ member of Eq.~\eqref{eq:Rpoly} and gives $\alpha(1-u)/u$. Summing the two contributions, we obtain
\begin{equation}
\Rc\big[\Delta_{\rm marg}\big]=\frac{\alpha}{u}
\simeq\frac{4\alpha N}{3} ,
\label{eq:Rmarg}
\end{equation}
so that Eq.~\eqref{eq:nonremovable} with $k=1$ gives $\Delta n_s\simeq\Rc/N=\tfrac43\alpha$, which matches the order reported in Refs.~\cite{CodelloEtAl2015,Yuennan2025}. This means that the marginal deformation is not wholly non-removable, and the field-linear remainder displaces $n_s$.

The lower block of Table~\ref{tab:test} places these results on the same footing as the channels generated by the flow. We find that every proposal examined that raises $n_s$ to the ACT preferred value fails the removal test, and the required amplitude is dictated by the rate at which it fails. We do not read this as an argument against those proposals. It is a statement about the origin of their effect. The final column gives the amplitude at which a linear extrapolation of $\Delta n_s$ reaches the ACT central value, which we denote by $\Amp_{\rm ACT}$. It is a diagnostic and not a proposed amplitude, and it is meaningful only where it stays small.

The amplitudes required to reproduce the ACT central value span three orders of magnitude across the literature proposals, from $6.7\times10^{-5}$ for an $R^{3}$ term to $9.0\times10^{-2}$ for a matter logarithm. Evaluated at those amplitudes, the criterion gives the narrow band $\Rc(\phi_0)=0.21$, $0.43$, $0.57$, and $0.80$ for the $R^{3}$, polynomial, marginal, and matter-logarithm entries, respectively. Interestingly, this band is explained by Eq.~\eqref{eq:nonremovable}. With $N_*\Delta n_s=0.51$, it predicts $\Rc=\tfrac38\times0.51=0.19$ for $R^{3}$, where $k=2$, and $\Rc=0.51$ for the marginal case, where $k=1$, in agreement with the numerical values at the ten percent level expected of a leading-order estimate. On the other hand, the class-R entries, and the $c\,u$ diagnostic, reach the same displacement only at $|\Amp_{\rm ACT}|>3$, or equivalently at $|\Rc(\phi_0)|=5.2$ and $9.1$ for the $c\,u$ and $K\ell$ entries, respectively, which is an order of magnitude larger and far outside the perturbative regime. These channels therefore cannot supply the ACT displacement while remaining small corrections. Note that removable deformations can raise $n_s$, for $K<0$ or for the $c\,u$ entry with $c>0$. What they cannot do is reach the ACT displacement within the perturbative domain $|\Amp|\ll1$. Where a slow-roll solution exists at those amplitudes, the numerical shift falls short of the ACT value by a factor between three and ten, and hence the response saturates. For a given displacement of $n_s$, the relevant quantity is $\Rc$ at the reference scale, and the amplitude of the deformation follows from it.

\begin{table}[t]
\centering
\small
\begin{tabular}{lccccc}
\toprule
deformation $\Delta$ & $\Rc[\Delta]$ & class &
$\Amp$ & $\Delta n_s$ & $\Amp_{\rm ACT}$\\
\midrule
\multicolumn{6}{l}{\emph{generated by the flow, Eq.~\eqref{eq:Delta}, with $\mu=H$}}\\
$K\,\ell$ & $K/2$ & R &
 $10^{-2}$ & $-5.09\times10^{-6}$ & $-18$\\
$-b_G\,a\phi\,u\,\ell/(1-u)$ & $\ord(u\ln u)$ & S &
 $10^{-2}$ & $-7.00\times10^{-6}$ & $-13$\\
$c\,\ell^{2}$ & $c\,\ell=\ord(u)$ & S &
 $6\times10^{-4}$ & $-5.20\times10^{-7}$ & $-11$\\
$\hat B_\Lambda V_0\,\ell/\Vst$ & $\hat B_\Lambda(\tfrac12-\ell)/(1-u)^{2}$ & R &
 $10^{-2}$ & $+2.50\times10^{-5}$ & $+3.7$\\
$\hat\rho_\Lambda V_0/\Vst$ & $-\hat\rho_\Lambda/(1-u)^{2}$ & R &
 $10^{-2}$ & $-1.53\times10^{-5}$ & $-6.1$\\
\midrule
\multicolumn{6}{l}{\emph{the same flow with $\mu^{2}=R_{\rm J}/12$, Sec.~\ref{sec:scheme}}}\\
$K\,\ell_{\rm J}$ & $K/4u$ & N &
 $10^{-3}$ & $+3.28\times10^{-4}$ & $+2.8\times10^{-2}$\\
\midrule
\multicolumn{6}{l}{\emph{limits of the criterion}}\\
$\Vst(\phi+\phi_{\rm s})/\Vst-1$, shift & $-0.83\,\phi_{\rm s}/\mpl$ &
 exact\textsuperscript{a} & $10^{-1}$ & $0$ & none\\
$c\,u$ & $-c(1-u)/2$ & R &
 $2\times10^{-2}$ & $+1.78\times10^{-5}$ & $+10$\\
\midrule
\multicolumn{6}{l}{\emph{proposed in the literature}}\\
$R^{3}$ in $f(R)$ \cite{CurvACT,AntoniadisEtAl2026} &
 $-c(1-u)/2u^{2}$ & N &
 $10^{-8}$ & $-1.40\times10^{-6}$ & $-6.7\times10^{-5}$\\
marginal $R^{2(1-\alpha)}$ \cite{CodelloEtAl2015,Yuennan2025} &
 $\alpha/u$ & N &
 $10^{-3}$ & $+1.28\times10^{-3}$ & $+7.3\times10^{-3}$\\
$\delta\ln(\phi/\mpl)$ \cite{AttractorsRadACT} &
 $\delta(1-u)/2au\phi$ & N &
 $10^{-3}$ & $+1.04\times10^{-4}$ & $+9.0\times10^{-2}$\\
$c\,(a\phi)^{2}$, cf.~\cite{Ellis2025} &
 $c(1-u)a\phi/u$ & N &
 $10^{-3}$ & $+7.31\times10^{-3}$ & $+1.3\times10^{-3}$\\
$\Boxop^{-1}R$
 \cite{NojiriOdintsovOikonomou2020,NojiriOdintsovOikonomou2026} &
 not a function of $\Vst$ & --- &
 --- & --- & ---\\
\bottomrule
\end{tabular}
\caption{The criterion \eqref{eq:Rtest} applied to deformations of the
Starobinsky plateau. Column two lists $\Rc$ on the plateau, evaluated at the pivot scale $u_0=1.264\times10^{-2}$ where a number is quoted. Column three gives the class of Sec.~\ref{sec:iff}: removable at $\ord(u)$ (R), subleading (S), or not removable (N). Columns four and five record the
amplitude $\Amp$ at which the deformation was evaluated and the numerical shift of $n_s$
at $N_*=55$ from Appendix~\ref{app:num}. The last column is the amplitude at
which a linear extrapolation of $\Delta n_s$ reaches the ACT central value
$+9.3\times10^{-3}$. The
response is linear to about three percent at $\Amp=10^{-2}$ for the entries generated by the flow. In the literature block the deviation from linearity at the quoted amplitude is $6\%$ for the polynomial entry, $3\%$ for the marginal one and below $1\%$ for the others, so $\Amp_{\rm ACT}$ is approximate.
Every entry generated by the flow at $\mu=H$ has $|\Amp_{\rm ACT}|>3$, whereas the non-removable entries have $|\Amp_{\rm ACT}|<0.1$. The polynomial entry is a representative of corrections with field-dependent masses and is not a transcription of the potentials of Ref.~\cite{Ellis2025}. \textsuperscript{a}The $c_3$ direction of Eq.~\eqref{eq:invclass}. Its
$\Rc$ is constant to $\ord(u)$ and its residual vanishes identically.}
\label{tab:test}
\end{table}

\subsection{The improvement scale and other renormalization-group treatments}
\label{sec:scheme}

The scale choice determines whether the logarithm tracks the potential or a quantity that varies exponentially with the field, and the classification of the gravitational flow changes accordingly. The general statement follows from Eq.~\eqref{eq:nonremovable}. Let the improvement scale behave on the plateau as $\mu\propto u^{-s}$. Then $\ell\simeq s\,a(\phi-\phi_0)$ is linear in the field, and for $\Delta=K\ell$ one obtains
\begin{equation}
\Rc[K\ell]\simeq\frac{sK}{2u}\simeq\frac{2sKN}{3},
\qquad
\Delta n_s\simeq\frac{\Rc}{N}=\frac{2s}{3}\,K .
\label{eq:scalerule}
\end{equation}
The Einstein-frame Hubble rate has $s=0$, since $H\to\Ms/2$ on the plateau. This is the reason for Eq.~\eqref{eq:ell} and for the removability of the flow in Sec.~\ref{sec:results}.

On the other hand, the Jordan-frame curvature has $s=\tfrac12$. By using $R_{\rm J}=3\Ms^{2}(1-u)/u$ from Eq.~\eqref{eq:fRlemma} and $\mu^{2}=R_{\rm J}/12$, we obtain
\begin{equation}
\ell_{\rm J}=\frac12\ln\frac{(1-u)\,u_0}{(1-u_0)\,u}
=\frac12\left[\ln\frac{1-u}{1-u_0}+a(\phi-\phi_0)\right],
\qquad
\Rc[K\ell_{\rm J}]=\frac{K}{4u},
\label{eq:ellJ}
\end{equation}
so that $\Delta n_s\simeq K/3$. The numerical solution gives $\Delta n_s=+0.329\,K$ at $N_*=55$ in the linear regime. Here the deformation $K\ell_{\rm J}$ is the marginal deformation of Eq.~\eqref{eq:Rmarg} with $\alpha=K/4$. This is the result of Refs.~\cite{BenDayan2014,CodelloEtAl2015,LiuProkopecStarobinsky2018}, in which the logarithmic running of the $R^{2}$ coefficient is evaluated at the Jordan-frame curvature. More importantly, with this scale the vacuum-energy channel is no longer suppressed. The numerical solution gives $\Delta n_s=+9.3\times10^{-3}$ at $\hat B_\Lambda=0.032$, with $r$ larger than its tree-level value by $76\%$, and $\Delta n_s=+2.0\times10^{-2}$ at the upper edge $\hat B_\Lambda=0.141$ of the window in Eq.~\eqref{eq:window_L}. The estimate of Eq.~\eqref{eq:maxshift} therefore holds for the Einstein-frame scale and not for the Jordan-frame one. The mass and Planck-mass channels remain negligible in either case, since $K\le1.3\times10^{-6}$ there.

\begin{figure}[t]
\centering
\includegraphics[width=0.62\textwidth]{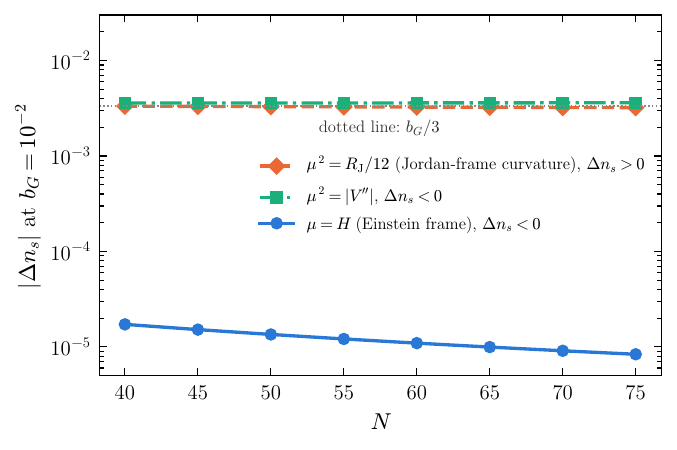}
\caption{$|\Delta n_s|$ in the $b_G$ channel at the illustrative value $b_G=10^{-2}$ for three improvement
prescriptions. The Jordan-frame curvature scale and $\mu^{2}=|V''|$ give shifts close to $\pm b_G/3$, as predicted by Eq.~\eqref{eq:scalerule}, whereas the Einstein-frame Hubble rate gives a shift smaller by a factor of about $300$.}
\label{fig:scheme}
\end{figure}

A third choice, $\mu^{2}=|V''|$, has $s=-\tfrac12$, since $|V''|\propto u$ on the plateau. It makes the logarithm $\ell_B\simeq-a\phi/2$, and the expansion parameter becomes $b_G(1-u)/4u=\ord(0.2)$ at the reference scale for $b_G=10^{-2}$. In addition, the improved potential develops a local maximum at $u_{\rm max}=\ord(b_G)$ in this truncated prescription, whose interpretation would require control of the omitted terms. The result is $\Delta n_s=-3.6\times10^{-3}$, which is close to $-b_G/3$, against $-1.2\times10^{-5}$ for $\mu=H$, as illustrated in Fig.~\ref{fig:scheme}. The figure shows the sensitivity of the truncation to the prescription. It does not establish the accuracy of any one prescription, and a larger shift or a local maximum is not by itself a reason to discard one. A rescaling $\mu=cH$ shifts $\ell$ by a constant and does not change the class. It amounts to a redefinition of the reference couplings, whose effect on the estimate is given in Sec.~\ref{sec:matching}.

It is worth noting that the three prescriptions are not related by a change of renormalization scheme. They resum, or in the present case parametrize, logarithms of different quantities. The question of which scale is appropriate is the same as the one encountered for the radiative corrections to Higgs inflation, where prescriptions defined in the Jordan and in the Einstein frame lead to different effective potentials \cite{BezrukovShaposhnikov2009,GeorgeMooijPostma2014}, although the one-loop divergences of the two frames agree on shell \cite{KamenshchikSteinwachs2015}. In this work, our choice follows from the definition of the completion. In class~E, the action in Eq.~\eqref{eq:SNL} is built on the Einstein-frame fields with a constant $\Lstar$, and the IR scale of the gravitational sector on the slow-roll background is the Einstein-frame Hubble rate. For a class-J completion, defined with a constant nonlocality scale in the Jordan frame, the Jordan-frame curvature is the corresponding scale, and the flow is then not removable. Removability of the gravitational running is therefore a property of class-E completions with the scale so identified. A derivation of the logarithm from the finite part of the one-loop effective action on the quasi-de Sitter background, which would settle its argument, is beyond the present truncation. We refer the readers to Ref.~\cite{BonannoDialektopoulosZarikas2024} for the identification of the scale within a gravitational action at the Lagrangian level.

In addition, the $\MSbar$ running of Lagrangian couplings employed here differs from the comoving-scale flow of primordial spectra developed for quadratic gravity with fakeons \cite{Anselmi2020a,Anselmi2020b} and from the functional RG improvements of the asymptotic-safety program \cite{BonannoReuter2002,BonannoSaueressig2015}. It also differs from the quadratic-gravity scenario of Ref.~\cite{LiuQuintinAfshordi2026}, where the one-loop running drives the slow-roll dynamics toward the IR and a large matter multiplicity within the beta functions displaces the predictions. That scenario operates within local asymptotically free quadratic gravity, whereas the present analysis concerns a super-renormalizable completion improved at $\mu=H$. The two analyses address different theories with different scale identifications, and their conclusions are not in conflict. Moreover, the Starobinsky model was mapped onto a segment of an asymptotically free quadratic-gravity trajectory in Ref.~\cite{PercacciVacca2025}. That mapping classifies the model within theory space and does not address how its predictions respond to the flow.

\subsection{Assumptions and limitations}
\label{sec:limits}

Before closing this section, it is worth summarizing the restrictions that define the scope of the present analysis:
\begin{enumerate}
\item \emph{Truncation.} The six-coupling system is a projection (see Sec.~\ref{sec:rg}). The improvement is a first-order substitution of running parameters into the tree-level potential, and the finite nonlocal parts of the one-loop effective action are omitted without an estimate.
\item \emph{Scale.} The improvement scale is the Einstein-frame Hubble rate. For the Jordan-frame curvature, the gravitational running is not removable (see Sec.~\ref{sec:scheme}).
\item \emph{Frame normalization.} The running Einstein--Hilbert coefficient is treated at first order with $Z=1$ and $\sigma=1$, and the $b_G$ entries of the tables refer to the substitution in Eq.~\eqref{eq:Delta}, which agrees with this choice within $3\%$ (see Sec.~\ref{sec:framenorm}).
\item \emph{Matching.} The coefficients of Eq.~\eqref{eq:BGgen} and the relation in Eq.~\eqref{eq:gammaMfix} are established for class~J. They are used for class~E as a dimensional estimate and as an assumption, respectively. In addition, the factors $\kappa_G$ and $\kappa_\Lambda$ depend on the zeros of the asymptotic polynomial through Eq.~\eqref{eq:kappas}, and the window in Eq.~\eqref{eq:window_L} scales as $|\kappa_\Lambda|^{-1/4}$.
\item \emph{Decoupling and renormalization condition.} The minimal-subtraction coefficients generated at the scale $\Lstar$ are used at $\mu=H_*<\Lstar$, and the estimate in Eq.~\eqref{eq:maxshift} depends on the condition imposed on the vacuum energy at the pivot scale (see Sec.~\ref{sec:matching}). Moreover, the vacuum-energy term is trusted on the plateau only.
\item \emph{Post-inflationary history.} The $e$-fold estimate of Sec.~\ref{sec:analytic} assumes the gravitational decay channel and matter-dominated reheating.
\item \emph{Poles.} The estimates of Sec.~\ref{sec:poles} concern loop-induced zeros of the flat-space propagator on the real momentum axis. They have not been verified on the de Sitter background, where infinite-derivative theories carry additional complex-mass states at tree level \cite{Tokareva2024,FengYang2026}, which are classically stable provided ${\rm Im}(m^{2})^{2}<9H^{2}{\rm Re}(m^{2})$ \cite{Addazi2024}.
\item \emph{Slow-roll order.} The numerical results are high-precision evaluations of the first-order potential slow-roll system, and numerical precision is not physical accuracy.
\end{enumerate}
Concerning the last item, we have compared our results with the exact background evolution, which is obtained by integrating the Klein--Gordon and Friedmann equations up to $\eps_H=1$ and combined with the second-order spectral formula \cite{StewartLyth1993} in the Hubble-flow parameters (see Appendix~\ref{app:num} and Table~\ref{tab:background}). We find that the linear-response coefficients of $\Delta n_s$ change by $+5\%$ for $\gamma_M$, $+0.1\%$ for $b_G$, $+17\%$ for $\hat B_\Lambda$, $+13\%$ for $\hat\rho_\Lambda$ and $-0.6\%$ for the marginal deformation. The vacuum-energy channels are the most sensitive, because their deformation is largest near the end of inflation. However, none of these changes alters a classification or an order of magnitude. The mode equations have not been solved.

These restrictions do not affect the criterion as a statement about a given deformation, which is a property of its functional form. They affect the question of which deformation the gravitational running produces, and they bound the applicability of the numerical estimates and of the pole analysis.

\section{Conclusions}
\label{sec:concl}

In the present work, we have derived the first-order response of the Starobinsky predictions to a fractional deformation $\Delta$ of the Einstein-frame potential, see Eq.~\eqref{eq:response}. We have shown that it depends on $\Delta$ through $\Rc[\Delta]=(\Vst/\Vst')\Delta'$ and that, on the plateau, the shift of $\ln\eps$ is twice the logarithmic derivative of the trajectory average of $\Rc$. A deformation with constant $\Rc$ changes the E-model exponent and shifts $n_s$ at $\ord(\ln N/N^{2})$ only, whereas $\Rc\propto N^{k}$ shifts it by $[2k^{2}/(k+1)]\Rc/N$. The salient feature of this test is that it requires one differentiation and no computation of observables, and it accounts for the scaling of every entry of Table~\ref{tab:test}. In so doing, we turn the known fact that the attractor predictions depend only on the leading approach to the plateau into a quantitative statement about a multiplicative deformation of the canonical potential. These statements do not depend on a completion.

\medskip\noindent
Within a projected truncation of the local couplings and a first-order improvement at the Einstein-frame Hubble rate, we have demonstrated that the deformation induced by the one-loop running of $\mpl^{2}$, $\Ms^{2}$ and the vacuum energy agrees at first order with a change of the E-model exponent, i.e., $p=2+K$, because the RG time equals one half of $\ln\Vst$. This is a first-order correspondence and not an exact trajectory. The residual shifts are $\ord(\ln N/N^{2})$ in $n_s$. They are obtained in closed form for the constant-$\Rc$ direction in Eq.~\eqref{eq:dnsanalytic}, and with the leading logarithms of Eq.~\eqref{eq:asymptotic} for the four channels. With a consistent frame normalization, the running of $\mpl^{2}$ acts through the kinetic term and through the dimensionless plateau height $1/16\alpha_R$, and its effect is of the same class and size. However, neither the number of $e$-folds nor the tensor sector is protected. For the minimal gravitational decay channel, the variation of the matching of the pivot scale opposes the shape shift in all four channels, by $11\%$ in the direction $\gamma_M=b_G$ and by $14\%$ for the vacuum-energy running. In class~J, both $r$ and $n_T$ change by relative amounts of order $\delta\ln\eps=\ord(K\ln N/N)$, i.e., at most a few percent, while in class~E the tilt changes by a relative amount below $10^{-11}$.

\medskip\noindent
We have found that the coefficients of the running are sums over the zeros of the asymptotic polynomial of the form factor, see Eq.~\eqref{eq:BGmass}, where each zero contributes as a minimally coupled scalar. Real zeros give $\hat B_\Lambda>0$, which raises $n_s$, and positive real zeros give $b_G>0$, which lowers it by a negligible amount. In addition, a monomial gives no running. Under a naturalness condition on the vacuum-energy running, and for vanishing vacuum energy at the pivot scale, the matching to the nonlocality scale gives $|\Delta n_s|$ between $2.6\times10^{-4}$ and $3.6\times10^{-4}$ depending on the evaluation. This is a factor of about $30$ below the displacement between the ACT central value and the tree-level prediction. The estimate becomes $5.8\times10^{-4}$ if a vacuum energy of the same size is allowed at the pivot scale. We notice, however, that it is a conditional estimate and not a bound on local gravitational corrections, on weakly nonlocal completions in general, or on other reheating histories. It depends on the renormalization condition for the vacuum energy, it uses minimal-subtraction coefficients below the nonlocality scale, and its translation into a prediction of a particular completion requires a consistent matching of the independent couplings and control of the finite quantum terms. Moreover, the matter contribution to the running of $\Ms^{2}$ follows from $\gamma_M=b_G-B_R/\alpha_R$, to which only nonminimally coupled scalars contribute. For the SM, it is fixed by the Higgs nonminimal coupling and amounts to $\Delta n_s=-1.2\times10^{-14}(\xi_H-1/6)^{2}$.

\medskip\noindent
More importantly, the conclusion that the gravitational running does not displace $n_s$ is conditional on the improvement scale. For a scale that varies as $u^{-s}$ on the plateau, the same flow gives $\Delta n_s\simeq\tfrac23sK$. The Jordan-frame curvature has $s=\tfrac12$ and reproduces the marginally deformed Starobinsky model, for which the vacuum-energy channel can reach the ACT displacement within the same window. Which scale applies is tied to the frame in which the completion is defined, and a calculation of the finite one-loop effective action on the inflationary background is needed to settle it.

\medskip\noindent
An $R^{n}$ correction has $\Rc=\ord(N^{n-1})$ and a marginal deformation has $\Rc=\ord(\alpha N)$. We have found that none of the proposals examined that raise $n_s$ is removable, and a removable deformation would need an amplitude $|K|$ of order ten to reach the ACT value. The amplitudes required span three orders of magnitude, while the value of $\Rc$ at the reference scale needed for that shift lies between $0.2$ and $0.8$ in every case, as predicted by Eq.~\eqref{eq:nonremovable}.

\medskip\noindent
There are several issues which are worth further investigation. The first one is the determination of the argument of the logarithm from the finite part of the one-loop effective action, in both classes of completions. Another one is a de Sitter treatment of the improved propagator that incorporates the background-induced states \cite{Tokareva2024,FengYang2026}. In addition, a calculation of the coefficients in Eq.~\eqref{eq:BGmass} for the class-E completion, the interplay between removable and non-removable corrections once matter fields are fully coupled, and a dedicated analysis of the reheating channel deserve further study. A solution of the mode equations for representative deformations would test the error estimates of Sec.~\ref{sec:limits}. For the primordial bispectrum, exact field shifts leave the squeezed limit unchanged, whereas the E-model exponent direction changes it through the single-field consistency relation by an amount proportional to the residual $\Delta n_s$ computed here. Whether removable deformations leave the equilateral configuration unchanged at leading order is an open question. The machinery required to evaluate the nonlocal three-point interactions is available \cite{KoshelevKumarMazumdarStarobinsky2020,NonGauss2022,KoshelevNaskar2025}. We leave these issues for future work.

\acknowledgments
We thank Boris Latosh for bringing Ref.~\cite{Arbuzov:2020pgp} to our attention and for useful comments. This research has received funding support from the NSRF via the Program Management Unit for Human Resources and Institutional Development, Research and Innovation [grant number B13F680083]. DS has received funding support from the Fundamental Fund of Khon Kaen University. Generative AI tools (Claude, Anthropic; ChatGPT, OpenAI; and Grok, xAI) were used to assist with drafting and revising the manuscript text. The authors directed this assistance, verified all scientific content and equations against their calculations, and take full responsibility for the paper.

\section*{Data availability}
There are no publicly available research data that support this manuscript.
Requests for further information should be sent to the authors.

\appendix

\section{Scalaron potential, slow roll and the \texorpdfstring{$N$}{N}-map}
\label{app:derivs}

In this appendix, we collect the derivatives of the scalaron potential, the slow-roll parameters and the $e$-fold map used in Secs.~\ref{sec:setup} and \ref{sec:results}, together with the running of the nonminimal coupling that justifies setting $\xi=0$. By using $u=e^{-a\phi}$, $a=\sqrt{2/3}/\mpl$, $V_0=\tfrac34\mpl^{2}\Ms^{2}$ and $\Vst=V_0(1-u)^{2}$, so that $u'=-au$ and $2V_0a^{2}=\Ms^{2}$, we obtain
\begin{equation}
\begin{aligned}
\Vst'&=\sqrt{\tfrac32}\mpl\Ms^{2}u(1-u),
&\Vst''&=\Ms^{2}\left(2u^{2}-u\right),\\
\Vst'''&=\sqrt{\tfrac23}\frac{\Ms^{2}}{\mpl}\left(u-4u^{2}\right),
&\Vst''''&=\frac{2\Ms^{2}}{3\mpl^{2}}\left(8u^{2}-u\right),
\end{aligned}
\label{eq:Vderivs}
\end{equation}
where $\Vst''(0)=\Ms^{2}$ identifies the scalaron mass. With $\mpl^{2}a^{2}=2/3$, the slow-roll parameters read
\begin{equation}
\eps_V=\frac43\frac{u^{2}}{(1-u)^{2}},
\qquad
\eta_V=\frac43\frac{2u^{2}-u}{(1-u)^{2}} .
\label{eq:appsr}
\end{equation}
Inflation ends at $\eps_V=1$, i.e., $u_{\rm end}=(1+2/\sqrt3)^{-1}=2\sqrt3-3=0.4641$ and $\phi_{\rm end}=0.9402\mpl$. From $\dd N=\Vst\dd\phi/(\mpl^{2}\Vst')$ with $\dd\phi=-\dd u/(au)$, one obtains
\begin{equation}
\frac{\dd N}{\dd u}=-\frac34\frac{1-u}{u^{2}}
\qquad\Longrightarrow\qquad
N=\frac34\left(\frac1u+\ln u-c_{\rm end}\right),
\qquad
u_N\simeq\frac{3}{4N} ,
\label{eq:appN}
\end{equation}
with $c_{\rm end}=1/u_{\rm end}+\ln u_{\rm end}=1.387$, and the general-$p$ versions follow by $2\to p$. Along the trajectory, one has $\dd\ln\eps_V/\dd N=-8u/3(1-u)^{2}$, which equals $-2/u_{\rm end}$ at the end point. In addition, the nonminimal coupling runs as
\begin{equation}
\be_\xi(\phi_0)=\left(\xi-\frac16\right)\frac{V''''(\phi_0)}{(4\pi)^{2}}
=\left(\xi-\frac16\right)\frac{2\Ms^{2}}{3(4\pi)^{2}\mpl^{2}}
\left(8u_0^{2}-u_0\right),
\label{eq:betaxi}
\end{equation}
which is negligible for $\Ms^{2}/\mpl^{2}=1.5\times10^{-10}$.

\section{Heat-kernel conventions and the pole-to-beta rule}
\label{app:hk}

Here we record the heat-kernel form used in this work and the rule that maps a pole to a beta function, together with the check that fixes it. Eq.~\eqref{eq:scalarDiv} follows from the Seeley--DeWitt coefficient \cite{BarvinskyVilkovisky1985,Vassilevich2003} written for $\mathcal{D}=-\Boxop+\mathcal{E}$. It reads
\begin{equation}
a_2=\frac{1}{180}\left(R_{\mu\nu\rho\sigma}^{2}-R_{\mu\nu}^{2}\right)\mathbf 1
+\frac12\left(\mathcal{E}-\frac R6\mathbf 1\right)^{2}
+\frac{1}{30}\Boxop R\,\mathbf 1-\frac16\Boxop \mathcal{E}
+\frac{1}{12}\Omega_{\mu\nu}\Omega^{\mu\nu} .
\label{eq:a2}
\end{equation}
Here $\mathbf 1$ is the identity in field space and $\Omega_{\mu\nu}$ is the curvature of the connection. Note that the literature form is usually quoted for $-(\nabla^{2}+\mathcal{E}')$, and the substitution $\mathcal{E}'=-\mathcal{E}$ must be made in every term linear in $\mathcal{E}'$, which reverses the sign of the cross term as
\begin{equation}
\frac{180\mathcal{E}'^{2}+60R\mathcal{E}'+5R^{2}}{360}\bigg|_{\mathcal{E}'=-\mathcal{E}}
=\frac{180\mathcal{E}^{2}-60R\mathcal{E}+5R^{2}}{360}
=\frac12\left(\mathcal{E}-\frac R6\right)^{2}.
\label{eq:a2map}
\end{equation}
Omitting the cross term, or keeping it without reversing its sign, destroys the $(\xi-1/6)$ structure.

Minimal subtraction gives $\mu\,\dd\lambda_O/\dd\mu=-X$ when the effective action contains the term $\Gamma\supset+\int\sqrt{-g}\,\lambda_O O$ and its divergent part is $\Gamma_{\rm div}=+X/(2\bar\epsd)\int\sqrt{-g}\,O$. For the potential, the action term carries the opposite sign, and hence $\be_V=+X_V$. This is confirmed by the Coleman--Weinberg potential, i.e., $\be_V=-\mu\partial_\mu V_1=m_{\rm eff}^{4}/32\pi^{2}=[V'']^{2}/2(4\pi)^{2}$, where $V_1$ is the one-loop potential. For the $R^{2}$ term of Eq.~\eqref{eq:scalarDiv}, the rule gives $\be_{\alpha_R}=-(\xi-1/6)^{2}/2(4\pi)^{2}$ per real scalar, which is Eq.~\eqref{eq:BRmatter}. The source papers \cite{ModestoRachwalShapiro2018,RachwalModestoPinzulShapiro2021} define $\Gamma^{(1)}_{\rm div}\equiv-\mu^{n-4}(2\epsd)^{-1}\int\sqrt{|g|}\be_iO_i$ with $\epsd=(4-n)/2$ in $n$ dimensions. Their $\be_i$ are beta functions in our sense, i.e., $\mu\,\dd\omega_i/\dd\mu=\be_i$. Their Einstein--Hilbert and vacuum terms are $S_{\rm EH}=-(16\pi G)^{-1}\int\sqrt{|g|}\,(R+2\Lambda_{\rm cc})$, where $\Lambda_{\rm cc}$ is the cosmological constant, so that $\omega_{\rm EH}=-\mpl^{2}/2$ and $\omega_{\rm cc}=-\Lambda_{\rm cc}/8\pi G=-\rho_\Lambda$. This is the action $\int\sqrt{-g}\,[\tfrac12\mpl^{2}R-\rho_\Lambda]$ of our conventions with the opposite sign of the curvature scalar. Therefore, we obtain
\begin{equation}
B_G\equiv\mu\frac{\dd}{\dd\mu}\Big(\frac{\mpl^{2}}{2}\Big)=-\be_G,
\qquad
B_\Lambda\equiv\mu\frac{\dd\rho_\Lambda}{\dd\mu}=-\be_{\rm cc}=-B_{\omega_{\rm cc}} ,
\label{eq:signmap}
\end{equation}
which is Eq.~\eqref{eq:BGgen}. Our conventions for these two terms coincide with those of Ref.~\cite{ModestoOrlando2026}, which uses the signature $(-,+,+,+)$, a positive Einstein--Hilbert coefficient, the vacuum term $-\lambda$, the divergent part $\Gamma_{\rm div}=-[2(4\pi)^{2}\epsd]^{-1}\sum_i\be_i\int\sqrt{-g}\,O_i$ and the running of the couplings $\alpha_i(\mu)=\alpha_i(\mu_0)+\be_i(4\pi)^{-2}\ln(\mu/\mu_0)$. With $X=-\be_i/(4\pi)^{2}$, this is the rule stated above.

The dictionary is collected in Table~\ref{tab:signs}. Its first five rows follow from Eq.~\eqref{eq:scalarDiv} and the rule above, with the counterterm $\delta\lambda_O=-X/2\bar\epsd$ in each case, and the second row is the one checked against the Coleman--Weinberg potential. The entries for $R$ and $R\phi^{2}$ are field dependent, in line with the projection discussed in Sec.~\ref{sec:rg}. The last two rows are imported, and there are two checks which support their signs. Firstly, the relation $B_\Lambda=-B_{\omega_{\rm cc}}$ does not depend on the signature, because the unit operator carries no curvature sign and the combination $\tfrac12(\omega_{\Npol-1}/\omega_{\Npol})^{2}-\omega_{\Npol-2}/\omega_{\Npol}$ is even in $\Boxop$. Secondly, in terms of the zeros of the asymptotic polynomials, Eq.~\eqref{eq:BGgen} becomes Eq.~\eqref{eq:BGmass}, where we have used $\sum_im_i^{4}=(\sum_im_i^{2})^{2}-2\sum_{i<j}m_i^{2}m_j^{2}$. For one minimally coupled real scalar of mass $m$, rows one and three of the table give $\be_{\rho_\Lambda}=m^{4}/2(4\pi)^{2}$ and $\mu\,\dd(\tfrac12\mpl^{2})/\dd\mu=m^{2}/6(4\pi)^{2}$. This means that each zero contributes to $B_\Lambda$ and $B_G$ with the sign and the normalization of one such scalar per degree of freedom, i.e., five in the spin-two sector and one in the spin-zero sector, as expected for the functional determinant of a factorized operator. The identification $\omega_{\Npol-1}/\omega_{\Npol}=\sum_im_i^{2}$ has been checked against Ref.~\cite{ModestoRachwalShapiro2018} as follows. Firstly, the vacuum-energy divergence, Eq.~(34) of that reference, is fixed by the flat-space determinant of the factorized operator and does not depend on the sign convention of the d'Alembertian. It agrees with the scalar value only if the effective action of that reference carries the Lorentzian sign, with $\omega_{\rm cc}=-\rho_\Lambda$ as stated there. Secondly, the Einstein--Hilbert term with $\omega_{\rm EH}=-\mpl^{2}/2$ then requires a curvature scalar of the opposite sign to ours. For the signature $(+,-,-,-)$, which produces this sign, the d'Alembertian is also minus ours, the operator of a mode of mass $m$ is $\Boxop+m^{2}$, and the identification follows. Thirdly, with this identification the divergence linear in $R$, Eq.~(33) of that reference, agrees in sign and normalization with one minimally coupled scalar per degree of freedom, with the same relative factor of five between the tensor and scalar modes that is noted there for both divergences. Note that the nonlocal example in Sec.~4 of that reference is written with the Euclidean d'Alembertian, for which the ratio $\omega_{\Npol-1}/\omega_{\Npol}$ has the opposite sign in terms of the $m_i^{2}$. Since the signature is not stated explicitly in that reference, the sign of $B_G$ rests on these consistency arguments, whereas the sign of $B_\Lambda$ does not depend on them.

\begin{table}[t]
\centering\small\setlength{\tabcolsep}{4pt}
\begin{tabular}{llll}
\toprule
operator $O$ & $\lambda_O$ in $\Gamma$ & residue $X$ & beta function\\
\midrule
$1$ (vacuum, $m^{2}=V''_{\rm min}$) & $-\rho_\Lambda$ & $m^{4}/2(4\pi)^{2}$ & $\be_{\rho_\Lambda}=+m^{4}/2(4\pi)^{2}$\\
$1$ (potential) & $-V$ & $[V'']^{2}/2(4\pi)^{2}$ & $\be_V=+[V'']^{2}/2(4\pi)^{2}$\\
$R$ & $\tfrac12\mpl^{2}$ & $(\xi-\tfrac16)V''/(4\pi)^{2}$ & $\be_{\mpl^{2}}=-2(\xi-\tfrac16)V''/(4\pi)^{2}$\\
$R\phi^{2}$ & $-\tfrac12\xi$ & $(\xi-\tfrac16)V''''/2(4\pi)^{2}$ & $\be_\xi=+(\xi-\tfrac16)V''''/(4\pi)^{2}$\\
$R^{2}$ & $\alpha_R$ & $(\xi-\tfrac16)^{2}/2(4\pi)^{2}$ & $\be_{\alpha_R}=-(\xi-\tfrac16)^{2}/2(4\pi)^{2}$\\
\midrule
$R$, imported \cite{ModestoRachwalShapiro2018} & $\tfrac12\mpl^{2}=-\omega_{\rm EH}$ & & $B_G=-\be_G$, Eq.~\eqref{eq:BGmass}\\
$1$, imported \cite{ModestoRachwalShapiro2018} & $-\rho_\Lambda=\omega_{\rm cc}$ & & $B_\Lambda=-\be_{\rm cc}$, Eq.~\eqref{eq:BGmass}\\
\bottomrule
\end{tabular}
\caption{Sign dictionary for the one-loop running. For $\Gamma\supset+\int\sqrt{-g}\,\lambda_OO$ and $\Gamma_{\rm div}=+X/(2\bar\epsd)\int\sqrt{-g}\,O$ the rule is $\mu\,\dd\lambda_O/\dd\mu=-X$. The upper block is the contribution of one real scalar with $\mathcal{L}_\phi=-\tfrac12(\nabla\phi)^{2}-V-\tfrac12\xi R\phi^{2}$. The lower block gives the imported gravitational coefficients in the conventions of Ref.~\cite{ModestoOrlando2026}, with $m_{i,j}^{2}$ the zeros of the asymptotic polynomials defined above Eq.~\eqref{eq:BGmass}.}
\label{tab:signs}
\end{table}

\section{Inversion of the \texorpdfstring{$e$}{e}-fold map}
\label{app:inv}

The closed forms of Eqs.~\eqref{eq:nsp} and \eqref{eq:rp} require the $e$-fold map of the $p$-family to be inverted to subleading order. Measured from the end point in Eq.~\eqref{eq:pexact}, the $p$-family map is given by
\begin{equation}
N=\frac{3}{2p}\left[\frac1u+\ln u-\frac{p+\sqrt3}{\sqrt3}
-\ln\frac{\sqrt3}{p+\sqrt3}\right].
\label{eq:pNmapfull}
\end{equation}
Inverting with $u_N=x(1+A_1x\ln x+B_1x)$ and $x=3/(2pN)$, matching the coefficient of $\ln x$ fixes $A_1=1$, and the constant term fixes
\begin{equation}
B_1=\ln\left(p+\sqrt3\right)-\frac{p}{\sqrt3}-\frac12\ln3-1 .
\label{eq:B1}
\end{equation}
From Eq.~\eqref{eq:pexact}, one finds
\begin{equation}
\begin{aligned}
n_s-1&=-\frac{4pu+2p^{2}u^{2}}{3(1-u)^{2}}
=-\frac{4p}{3}u-\left(\frac{8p}{3}+\frac{2p^{2}}{3}\right)u^{2}+\ord(u^{3}),\\
r&=\frac{16p^{2}}{3}u^{2}\left(1+2u\right)+\ord(u^{4}),
\end{aligned}
\label{eq:nsru}
\end{equation}
and substituting $u_N$, we obtain
\begin{equation}
n_s-1=-\frac2N-\frac{1}{N^{2}}\left[\frac3p\left(\ln x+B_1\right)+\frac6p+\frac32\right],
\qquad
r=\frac{12}{N^{2}}\left[1+\frac{3}{pN}\left(\ln x+B_1+1\right)\right],
\label{eq:nsrN}
\end{equation}
with $\ln x+B_1=-\ln[2pN/\sqrt3(p+\sqrt3)]-p/\sqrt3-1$. These are Eqs.~\eqref{eq:nsp} and \eqref{eq:rp}. Note that the constant $-p/\sqrt3$ in $\ln x+B_1+1$ is of the same order in $1/N$ as the logarithm and must be kept in $r$.

\section{First-order response in the \texorpdfstring{$N$}{N}-formalism}
\label{app:response}

In this appendix, we derive Eq.~\eqref{eq:response}, give its explicit form on the Starobinsky trajectory, and record its numerical validation.

\noindent
We work at first order in slow roll, $\dd N=V\,\dd\phi/(\mpl^{2}V')$ with $N$ counted from the end of inflation at $\eps_V=1$, and at first order in the deformation $\Delta$ of $V=\Vst[1+\Delta]$. The deformation may depend on the field value $\phi_0$ at the pivot scale as a fixed parameter.

\noindent
With $\Rc$ defined in Eq.~\eqref{eq:Rdef}, one has
\begin{equation}
\frac{V'}{V}=\frac{\Vst'}{\Vst}+\frac{\Delta'}{1+\Delta}
=\frac{\Vst'}{\Vst}\left[1+\Rc\right]+\ord(\Delta^{2}) .
\label{eq:app_lnV}
\end{equation}
Hence, at a given field value, we obtain
\begin{equation}
\eps_V(\phi)=\eps_0(\phi)\left[1+2\Rc(\phi)\right],
\qquad
\frac{\dd N}{\dd\phi}=\frac{\dd N_0}{\dd\phi}\left[1-\Rc(\phi)\right],
\label{eq:app_local}
\end{equation}
where $\eps_0$ and $N_0$ refer to $\Vst$. The end point moves to $\bar\phi_{\rm end}$ with $\eps_0(\bar\phi_{\rm end})[1+2\Rc_{\rm end}]=1$, which in terms of the tree-level $e$-fold variable reads
\begin{equation}
N_0(\bar\phi_{\rm end})=-\frac{2\Rc_{\rm end}}{(\dd\ln\eps_0/\dd N)_{\rm end}}=u_{\rm end}\,\Rc_{\rm end} ,
\label{eq:app_end}
\end{equation}
where we have used $(\dd\ln\eps_0/\dd N)_{\rm end}=-2/u_{\rm end}$ from Appendix~\ref{app:derivs}. Integrating the second relation of Eq.~\eqref{eq:app_local} from $\bar\phi_{\rm end}$, one obtains
\begin{equation}
N(\phi)=N_0(\phi)-N_0(\bar\phi_{\rm end})-\int_{0}^{N_0(\phi)}\Rc\,\dd N_0 ,
\label{eq:app_Nmap}
\end{equation}
and hence the field value $\phi_N$ reached $N$ $e$-folds before the end satisfies $N_0(\phi_N)=N+I(N)$ with $I(N)=\int_0^{N}\Rc\,\dd N'+u_{\rm end}\Rc_{\rm end}$ at first order. Substituting this into the first relation of Eq.~\eqref{eq:app_local}, i.e., $\eps(N)=\eps_0(N+I)[1+2\Rc(N)]$, gives the first equation of \eqref{eq:response}. The second and third ones follow from $r=16\eps$ and from Eq.~\eqref{eq:Nformalism}, which is an identity for $n_s=1-6\eps_V+2\eta_V$ when $N$ is defined by the slow-roll map. The derivative with respect to $N$ is taken along the trajectory of a fixed potential, and therefore $\phi_0$ is held fixed and set equal to $\phi_N$ afterwards.

\noindent
In terms of $u$, with $\dd N_0=-\tfrac34(1-u)u^{-2}\dd u$, the response takes the form
\begin{equation}
\begin{aligned}
\delta\ln\eps&=2\,\Rc(u)-\frac{2u}{(1-u)^{2}}
\left[\int_{u}^{u_{\rm end}}\Rc(x)\,\frac{1-x}{x^{2}}\,\dd x+\frac43u_{\rm end}\Rc(u_{\rm end})\right],\\
\Delta n_s&=-\frac{8u^{2}}{3(1-u)^{2}}\,\delta\ln\eps-\frac{4u^{2}}{3(1-u)}\,\partial_u\delta\ln\eps\Big|_{u_0} .
\end{aligned}
\label{eq:app_u}
\end{equation}
For constant $\Rc=K/2$, the integral is elementary and one obtains
\begin{equation}
\delta\ln\eps=K\left[1-\frac{1+u\ln u-\left(c_{\rm end}-\tfrac43u_{\rm end}\right)u}{(1-u)^{2}}\right]
=-K\,u\left[\ln u+2-c_{\rm end}+\tfrac43u_{\rm end}\right]+\ord(u^{2}\ln u) ,
\label{eq:app_const}
\end{equation}
with $c_{\rm end}-\tfrac43u_{\rm end}=5-2\sqrt3+\ln(2\sqrt3-3)=0.768$. With $u=3/4N$, this is the $\Delta r/r$ of Eq.~\eqref{eq:dnsanalytic}, and the derivative gives $\Delta n_s=\tfrac83\cdot\tfrac K2\,u^{2}[\ln u+3-c_{\rm end}+\tfrac43u_{\rm end}]$, which is the $\Delta n_s^{\rm shape}$ of the same equation. For $\Rc=w_0+w_1u+\rho_Lu\ln u$, the additional terms are given by
\begin{equation}
N^{2}\Delta n_s\supset-\frac34\left[w_L\ln^{2}u+2\left(w_1+2w_L\right)\ln u+4w_1+2w_L-2\tilde C\right],
\label{eq:app_logs}
\end{equation}
with $u=3/4N$ and $\tilde C$ being a constant that depends on $\Rc$ over the whole trajectory. The vacuum-energy channels have $(w_0,w_1,w_L)=(-1,-2,0)$ per unit $\hat\rho_\Lambda$ and $(\tfrac12,2,0)$ per unit $\hat B_\Lambda$, which gives the coefficients $-\tfrac32$ and $+\tfrac94$ of $\ln N$ in Eq.~\eqref{eq:asymptotic}, respectively. For $\hat\rho_\Lambda$, the constant is also elementary, i.e., $\tilde C=-2\ln u_{\rm end}+\ln(1-u_{\rm end})+\tfrac43u_{\rm end}[1-(1-u_{\rm end})^{-2}]=-0.624$, and reproduces $1.285$. The exponent shift of the $b_G$ channel has $\Rc=-\tfrac12u_0(\ln u+1)+u\ln u+\tfrac12u$ at leading order, where the terms proportional to $u_0$ contribute to $\Delta n_s$ only at $\ord(\ln^{2}N/N^{3})$, and $(w_1,w_L)=(\tfrac12,1)$ gives $-\tfrac34\ln^{2}N+\tfrac34(5-2\ln\tfrac43)\ln N$.

\noindent
For a kinetic function $\mathcal{K}=1+\delta_{\mathcal K}$ in front of $\tfrac12(\nabla\phi)^{2}$, as in Eq.~\eqref{eq:einsteinframe}, the slow-roll relations are $\eps=(\mpl^{2}/2\mathcal{K})(V'/V)^{2}$ and $\dd N=\mathcal{K}V\dd\phi/(\mpl^{2}V')$. At first order, one has $\eps=\eps_0(1+2\Rc-\delta_{\mathcal K})$ and $\dd N=\dd N_0(1-\Rc+\delta_{\mathcal K})$ at fixed field, and the steps above give
\begin{equation}
\delta\ln\eps(N)=\left(2\Rc-\delta_{\mathcal K}\right)(N)+\frac{\dd\ln\eps_0}{\dd N}
\left[\int_0^{N}\left(\Rc-\delta_{\mathcal K}\right)\dd N'+\frac{u_{\rm end}}{2}\left(2\Rc-\delta_{\mathcal K}\right)_{\rm end}\right] ,
\label{eq:app_kinetic}
\end{equation}
which reduces to Eq.~\eqref{eq:response} for $\delta_{\mathcal K}=0$. For $\delta_{\mathcal K}=-\ell$ and $\Rc=0$, it gives $\Delta n_s=-9.973\times10^{-4}$ at $N=55$, which is equal to the numerical solution of the non-canonical system to the digits quoted. This is the coefficient $c_{\mathcal K}$ of Eq.~\eqref{eq:frameresult} for $\sigma=0$.

\noindent
We close this section with the consistency checks of the results, and they are as follows. (i) For constant $\Rc$, Eq.~\eqref{eq:app_const} agrees with the direct expansion of the E-model family in Appendix~\ref{app:inv}. (ii) Evaluated by quadrature, Eq.~\eqref{eq:app_u} agrees with the numerical solution of Appendix~\ref{app:num} at amplitude $10^{-6}$ to a relative accuracy of $10^{-6}$ in all four channels at $N=40$, $55$ and $75$, for both $\Delta n_s$ and $\Delta r/r$. (iii) For $\Rc\propto N^{k}$, it reproduces the numerical shifts of the non-removable entries of Table~\ref{tab:test} at the ten percent level, as expected at leading order in $1/N$, i.e., $-1.50\times10^{-6}$ against $-1.40\times10^{-6}$ for the $R^{3}$ entry and $+1.44\times10^{-3}$ against $+1.28\times10^{-3}$ for the marginal one.

\section{Accuracy of the slow-roll evaluation}
\label{app:num}

The results labeled numerical in this work are evaluations of the first-order potential slow-roll system without a large-$N$ expansion. The improved potential is built from Eq.~\eqref{eq:Delta} with $\ell$ given by Eq.~\eqref{eq:ell}, the end point follows from $\eps_V=1$, and the $e$-fold integral is evaluated by adaptive quadrature. Since $\ell$ depends on $u_0$ while the pivot scale is defined by $\phi_0=\phi_N$, the procedure is iterated to self-consistency. In addition, all tabulated values were reproduced with a second, independent implementation.

In order to assess the accuracy of this evaluation, we integrated the exact background equation, $\dd^{2}\phi/\dd N^{2}-(3-\eps_1)(\dd\phi/\dd N-\mpl^{2}V'/V)=0$ with $\eps_1\equiv\eps_H=(\dd\phi/\dd N)^{2}/2\mpl^{2}$, from the slow-roll attractor to $\eps_1=1$ with the same self-consistent determination of $\phi_0$. The Hubble-flow parameters $\eps_{i+1}=-\dd\ln\eps_i/\dd N$ at $N_*=55$ were inserted in the first-order formula $n_s-1=-2\eps_1-\eps_2$ and in the second-order formula \cite{StewartLyth1993}, which reads
\begin{equation}
n_s-1=-2\eps_1-\eps_2-2\eps_1^{2}-(2\mathcal{C}+3)\eps_1\eps_2-\mathcal{C}\eps_2\eps_3,
\qquad
r=16\eps_1\left(1+\mathcal{C}\eps_2\right),
\qquad
\mathcal{C}=\gamma_E+\ln2-2 .
\label{eq:secondorder}
\end{equation}
In Table~\ref{tab:background}, we compare the resulting linear-response coefficients, obtained from central differences at amplitude $10^{-3}$, with those of the potential slow-roll system. Note that the mode equations were not solved.

\begin{table}[t]
\centering\small
\begin{tabular}{lcccc}
\toprule
 & potential SR & background, 1st order & background, 2nd order & change\\
\midrule
tree $n_s$ & $0.964977$ & $0.964445$ & $0.964885$ & \\
tree $r$ & $3.498\times10^{-3}$ & $3.643\times10^{-3}$ & $3.550\times10^{-3}$ & \\
\midrule
$\Delta n_s/\gamma_M$ & $-5.150\times10^{-4}$ & $-5.492\times10^{-4}$ & $-5.390\times10^{-4}$ & $+4.7\%$\\
$\Delta n_s/b_G$ & $-1.214\times10^{-3}$ & $-1.229\times10^{-3}$ & $-1.215\times10^{-3}$ & $+0.1\%$\\
$\Delta n_s/\hat B_\Lambda$ & $+2.578\times10^{-3}$ & $+3.074\times10^{-3}$ & $+3.009\times10^{-3}$ & $+16.7\%$\\
$\Delta n_s/\hat\rho_\Lambda$ & $-1.513\times10^{-3}$ & $-1.746\times10^{-3}$ & $-1.710\times10^{-3}$ & $+13.0\%$\\
$\Delta n_s/\alpha$ (marginal) & $+1.316$ & $+1.308$ & $+1.307$ & $-0.6\%$\\
\midrule
$\Delta n_s$ at $\hat B_\Lambda=0.141$ & $+2.63\times10^{-4}$ & $+2.92\times10^{-4}$ & $+2.86\times10^{-4}$ & $+9\%$\\
\bottomrule
\end{tabular}
\caption{Tree-level values and shifts at $N_*=55$ from the potential slow-roll (SR) system used in the main text and from the exact background evolution with first- and second-order spectral formulae. The last column is the change from the first to the third numerical column.}
\label{tab:background}
\end{table}

\ifjhep\bibliographystyle{JHEP}\else\IfFileExists{JHEP.bst}{\bibliographystyle{JHEP}}{\bibliographystyle{elsarticle-num}}\fi
\bibliography{refs}

\end{document}